\documentclass[12pt, draftclsnofoot, onecolumn]{IEEEtran}
\IEEEoverridecommandlockouts
\usepackage{etoolbox}
\newtoggle{2column}
\togglefalse{2column}
\usepackage{amsfonts}
\usepackage{amsmath}
\usepackage{balance}
\usepackage{subfigure}
\PassOptionsToPackage{hyphens}{url}
\usepackage{url}

\usepackage[pdftex]{graphicx}
\usepackage{epstopdf}
\usepackage{multirow}

\usepackage{pifont}
\usepackage{capt-of}

\usepackage{fancyhdr}
\newcommand{\Nr}{N_{\mathrm{r}}}
\newcommand{\Nt}{N_{\mathrm{t}}}
\newcommand{\Ns}{N_{\mathrm{s}}}
\newcommand{\NRF}{N_{\mathrm{RF}}}

\newcommand{\NQ}{N_{\mathrm{Q}}}
\newcommand{\Gr}{G_{\mathrm{r}}}
\newcommand{\Gt}{G_{\mathrm{t}}}
\newcommand{\Gc}{G_{\mathrm{c}}}
\newcommand{\SmBlc}{{\mathbf{S}}_{m}}
\newcommand{\SBlc}{{\mathbf{S}}_{1}}
\newcommand{\SMBlc}{{\mathbf{S}}_{M}}
\newcommand{\sone}{\bs_{m}[1]}

\newcommand{\sN}{\bs_{m}[N]}
\newcommand{\Nc}{N_{\mathrm{c}}}
\newcommand{\prc}{p_{\mathrm{rc}}}
\newcommand{\ar}{{\mathbf{a}}_{\mathrm{R}}}
\newcommand{\at}{{\mathbf{a}}_{\mathrm{T}}}
\newcommand{\atbar}{\bar{\mathbf{a}}_{\mathrm{T}}}
\newcommand{\AR}{{\mathbf{A}}_{\mathrm{R}}}
\newcommand{\AT}{{\mathbf{A}}_{\mathrm{T}}}
\newcommand{\ATbar}{\bar{\mathbf{A}}_{\mathrm{T}}}
\newcommand{\Arx}{{\mathbf{A}}_{\mathrm{rx}}}
\newcommand{\Atx}{{\mathbf{A}}_{\mathrm{tx}}}
\newcommand{\Atxbar}{\bar{\mathbf{A}}_{\mathrm{tx}}}
\newcommand{\jj}{{\mathrm{j}}}

\newcommand{\FRF}{{\mathbf{F}}_{\mathrm{RF}}}
\newcommand{\FBB}{{\mathbf{F}}_{\mathrm{BB}}}
\newcommand{\WRF}{{\mathbf{W}}_{\mathrm{RF}}}
\newcommand{\WBB}{{\mathbf{W}}_{\mathrm{BB}}}
\newcommand{\Exp}{{\mathbb{E}}}
\newcommand{\vect}{\mathrm{vec}}
\newcommand{\diag}{\mathrm{diag}}
\newcommand{\fbH}{\boldsymbol{H}}
\newcommand{\td}{\mathrm{td}}
\newcommand{\fd}{\mathrm{fd}}
\newcommand{\ttNMSE}{\mathtt{NMSE}}
\newcommand{\Ts}{T_\mathrm{s}}
\newcommand{\hchat}{\hat{\bh}_{\mathrm{c}}}

\newcommand{\be}{\begin{eqnarray}}
\newcommand{\ee}{\end{eqnarray}}

\newcommand{\figref}[1]{{Fig.}~\ref{#1}}
\newcommand{\secref}[1]{{Section}~\ref{#1}}

\def\Trans {{\rm T}}

\def\Ns{{N_\mathrm{s}}}
\def\Nt{{N_\mathrm{t}}}

\def\Nr{{N_\mathrm{r}}}

\def\Ns{{N_\mathrm{s}}}
\def\Np{{N_\mathrm{p}}}
\def\Nc{{N_\mathrm{c}}}

\def\No{{N_\mathrm{o}}}

\def\SNR{\mathrm{SNR}} 

\def\bb0{{\mathbb{0}}}

\def\bb{{\mathbf{b}}}

\def\bee{{\mathbf{e}}}

\def\bh{{\mathbf{h}}}

\def\bp{{\mathbf{p}}}

\def\br{{\mathbf{r}}}
\def\bs{{\mathbf{s}}}

\def\bv{{\mathbf{v}}}

\def\bx{{\mathbf{x}}}
\def\by{{\mathbf{y}}}

\def\bA{{\mathbf{A}}}
\def\bB{{\mathbf{B}}}
\def\bC{{\mathbf{C}}}

\def\bF{{\mathbf{F}}}

\def\bH{{\mathbf{H}}}
\def\bI{{\mathbf{I}}}

\def\bS{{\mathbf{S}}}

\def\bW{{\mathbf{W}}}

\def\bbC{{\mathbb{C}}}

\def\bbE{{\mathbb{E}}}

\def\bbR{{\mathbb{R}}}

\def\cA{\mathcal{A}}

\def\cN{\mathcal{N}}

\def\cS{\mathcal{S}}

\def\sfA{\mathsf{A}}

\def\sfD{\mathsf{D}}

\def\sf0{{\mathsf{0}}}

\def\bsf0{{\bm{\mathsf{0}}}}

\def\hc{\bh_{\mathrm{c}}}

\def\Gr{G_{\mathrm{r}}}
\def\Gt{G_{\mathrm{t}}}

\def\xi{x_{\mathrm{i}}}

\usepackage[usenames]{color}

\begin{document}
\hyphenation{multi-symbol}
\title{Channel Estimation for Hybrid Architecture Based Wideband Millimeter Wave Systems}

\author{ Kiran Venugopal, Ahmed Alkhateeb, Nuria Gonz\'{a}lez Prelcic,  and \\ Robert W. Heath, Jr.\\ 
	\thanks{This work was supported in part by the Intel-Verizon 5G research program and the National Science Foundation under Grant No. NSF-CCF-1319556. Part of the content of this work was submitted to ICASSP 2017 \cite{VenAlkPreHeath:Time-domain-chan-estimation-wideband-hybrid:16}. Kiran Venugopal, Ahmed Alkhateeb and Robert W. Heath, Jr. are with the University of Texas, Austin, TX, USA, and  Nuria Gonz\'{a}lez Prelcic is with the University of Vigo, Spain, Email: \tt{\{kiranv, aalkhateeb, rheath\}@utexas.edu, nuria@gts.uvigo.es}}
}
\date{} 
\maketitle

\vspace{-1cm}
\thispagestyle{empty}
\begin{abstract}
	Hybrid analog and digital precoding allows millimeter wave (mmWave) systems to achieve both array and multiplexing gain. The design of the hybrid precoders and combiners, though, is usually based on knowledge of the channel. Prior work on mmWave channel estimation with hybrid architectures focused on narrowband channels. Since mmWave systems will be wideband with frequency selectivity, it is vital to develop channel estimation solutions for hybrid architectures based  wideband mmWave systems. In this paper, we develop a sparse formulation and compressed sensing based solutions for the wideband mmWave channel estimation problem for hybrid architectures. First, we leverage the sparse structure of the frequency selective mmWave channels and formulate the channel estimation problem as a sparse recovery in both time and frequency domains. Then, we propose explicit channel estimation techniques for purely time or frequency domains and for combined time/frequency domains. Our solutions are suitable for both SC-FDE and OFDM systems. 
 Simulation results show that the proposed solutions achieve good channel estimation quality, while requiring small training overhead. Leveraging the hybrid architecture at the transceivers gives further improvement in estimation error performance and achievable rates.
 \end{abstract}


\section{Introduction}

Channel estimation in millimeter wave MIMO systems allows flexible design of hybrid analog/digital precoders and combiners under different optimization criteria.
Unfortunately, the hybrid constraint makes it challenging to directly estimate the channels, due to the presence of the analog beamforming / combining stage. Further, operating at mmWave frequencies  complicates the estimation of the channel
because the signal-to-noise-ratio (SNR) before beamforming is low and the dimensions of the channel matrices
associated with mmWave arrays \cite{HeathJr2016,Alkhateeb2014d} are large. While mmWave channel estimation has been
extensively studied in the last few years, most prior work assumed a narrowband channel model. Since mmWave systems are attractive due to their wide bandwidths, developing efficient mmWave channel estimation for
frequency selective channels is of great importance.

\subsection{Prior Work}
To avoid the explicit estimation of the channel, analog beam training solutions were proposed \cite{Wang2009,Chen2011,Hur2013}.  In beam training, the transmitter and receiver iteratively search for the beam pair that maximizes the link SNR \cite{Wang2009, Hur2013,Tsang2011}. This approach is used in IEEE standards like 802.11ad \cite{IEEE:11ad} and 802.15.3c \cite{15.3c:home}. The directional antenna patterns can be realized using a network of phase shifters. While analog beam training works for both narrowband and wideband systems, the downside is that the solution supports mainly a single communication stream, and the extensions to multi-stream and multi-user communication are non-trivial. Further, analog beamforming is normally subject to hardware constraints such as the quantization of the analog phase shifters, which limit their performance. Fully-digital architectures are the opposite to analog-only solutions, where every antennas is associated with an individual RF chain. This results however in high cost and power consumption at mmWave frequencies, making the fully-digital solutions unfeasible \cite{Singh2009,HeathJr2016,ElAyach2014}. 

To support multi-stream and multi-user transmissions in mmWave systems, hybrid analog / digital architectures were proposed \cite{ElAyach2014,Alkhateeb2013,Han2015,Mendez-Rial2016}. With hybrid architectures, the precoding / combining processing is divided between analog and digital domains. While hybrid architectures were shown to provide achievable rates close to those of fully-digital architectures \cite{ElAyach2014}, they pose more constraints that complicate the channel estimation problem \cite{HeathJr2016}. This is mainly because the channel is seen through the RF lens at the receiver baseband and because channel estimation has to be done before beamforming under low SNR conditions.

To address this problem in narrowband mmWave systems, \cite{Alkhateeb2014} proposed to formulate the hybrid precoding based mmWave channel estimation problem as a sparse recovery problem. This leverages the sparse structure of  mmWave channels in the angular domain making use of the {\em extended virtual channel model} \cite{HeathJr2016}.  In this approach, the MIMO channel is written in terms of dictionary matrices built from the transmit and receive steering vectors evaluated on a uniform grid of possible angles of arrival and departure (AoA/AoD). These dictionary matrices operate as a sparsifying basis for the channel matrix. Based on that, several channel estimation algorithms that use compressed sensing (CS) tools have been developed for hybrid architectures \cite{Alkhateeb2014,Mendez-Rial2016,Alkhateeb2015,Han2016a,Lee2014}, where the training/measurement matrices are designed using hybrid precoders and combiners. These techniques differ in the way these measurement matrices search for the dominant angles of arrival and departure. Solutions that make use of adaptive compressed sensing \cite{Alkhateeb2014, IweTew:Adaptive-strategies-for-target-detection:12, MalNow:Near-optimal-compressive-binary-search:12}, random compressed sensing \cite{RamVenMad:Compressive-adaptation-of-large:12,BerArmNix:Application-of-compressive-sensing:14, Alkhateeb2015, Han2016,Mendez-Rial2016}, joint random and adaptive compressed sensing \cite{Han2016a} were studied. Other non-compressed sensing techniques were also developed for mmWave channel estimation using subspace estimation \cite{Ghauch2015}, overlapped beams \cite{Kokshoorn2015}, and auxiliary beams \cite{Zhu2016}.

\subsection{Contributions} 
As mmWave systems will likely be wideband and frequency selective, developing wideband mmWave channel estimation techniques is crucial for practical mmWave systems. The aforementioned prior CS based solutions \cite{Alkhateeb2014,Alkhateeb2015,Han2016,Mendez-Rial2016,Ghauch2015,Kokshoorn2015,Zhu2016} focused on narrowband mmWave channels. Recently, wideband mmWave channel estimation using the hybrid architecture was considered in \cite{GaoHuDaiWan:Channel-estimation-mmWaveMassiveMIMO:16} for the first time, assuming an OFDM system and ideal settings. In this paper, building on our prior work in \cite{VenAlkPreHeath:Time-domain-chan-estimation-wideband-hybrid:16, VenAlkPreHeath:Exploiting-sparsity-wideband-mmWave-11ad-chanestimation:16}, we propose a novel mmWave channel estimation technique for wideband mmWave system, which works for both SC-FDE and OFDM systems. We also incorporate important system constraints like the frame structure, band limited nature of the pulse shaping filter used for the wideband system, and the hybrid architecture. One of the primary focuses of the paper is to reformulate the frequency selective channel estimation in large antenna systems as a sparse recovery problem, redefining the sparsifying dictionaries to account for the sparse nature of wideband frequency selective mmWave channels in both the angular and the delay domains. Once the channel is written in terms of the sparsifying dictionary matrices, the hardware constraints associated with the analog precoding stage are also introduced into the formulation of the channel estimation problem. With this key step, various algorithms available in CS literature 
can be used to fine tune the end performance. The main contributions can be summarized as follows:

\begin{itemize}
\item We define an appropriate sparsifying dictionary for frequency selective mmWave channels. This dictionary 
 depends on the transmit and receive array steering vectors evaluated on a uniform grid of possible AoAs/AoDs, and also on a raised cosine pulse shaping filter evaluated on a uniform grid of possible delays. This key step leads to a  representation  of the MIMO  channel matrix  that leverages the sparse structure of the mmWave channel in both the angular and delay domains. 
	\item We formulate the wideband mmWave channel estimation problem as a sparse recovery problem, in (1) the time domain, and in (2) the frequency domain. Important practical features critical for mmWave system modeling are incorporated in our sparse formulation. The proposed formulation simultaneously leverages the structure in the frequency selective large antenna mmWave channel and the frame structure assumed for data transmission. Unlike prior work which either relies on fully digital and/or OFDM systems for wideband channel estimation, our proposed approaches work both for SC-FDE and OFDM based frequency selective hybrid mmWave systems. 
	\item We propose explicit algorithms  to solve this sparse recovery problem in (1) purely time domain, (2) purely frequency domain, and (3) combined time-frequency domains. The different approaches proposed in this paper can be suitably used for different scenarios based on system level constraints and implementation. Our proposed time domain algorithms leverage the dictionary formulation that accounts for the sparsity in the delay domain, while the frequency domain techniques work independent of the delay domain sparsity constraints. 
	\item We leverage the hybrid architecture at the both at the transmitter and the receiver of mmWave wideband systems, unlike \cite{VenAlkPreHeath:Time-domain-chan-estimation-wideband-hybrid:16, VenAlkPreHeath:Exploiting-sparsity-wideband-mmWave-11ad-chanestimation:16}, to show how compressive sensing, hybrid precoding and combining result in low training overhead for explicit channel estimation in frequency selective mmWave systems. The proposed channel estimation techniques can be used to enable MIMO and multi-user communication in 802.11ad, as a potential application area. 
\end{itemize}

It is explained through simulation results that the proposed algorithms require significantly less training than when beam training (eg. IEEE 802.11ad) is used for estimating the dominant angles of arrival and departure of the channel. A strict comparison with existing beam training algorithms in terms of rate performance is not reasonable since they focus, not on estimating the explicit frequency selective mmWave MIMO channel, but on estimating beam pairs that give good link SNR. Ensuring low estimation error rates in our proposed algorithms, however, implies that efficient hybrid precoders and combiners can be designed to support rates similar to all-digital solutions \cite{AlkHea:Frequency-Selective-Hybrid:16}. We therefore rely mainly on the average error rates to compare the efficiency of our approaches.
We show that utilizing multiple RF chains at the transceivers further reduces the estimation error and the training overhead. Simulation results compare the three proposed techniques. The performance of the proposed techniques as system and channel parameters are varied are presented to identify which approach suits better for a given scenario. 

\textbf{Notation:} We use the following notation in the rest of the paper: bold uppercase $\mathbf{A}$ is used to denote matrices, bold lower case $\mathbf{a}$ denotes a column vector, and non-bold lower case $a$ is used to denote scalar values. We use $\mathcal{A}$ to denote a set. Further, $||\mathbf{A}||_F$ is the Frobenius norm, and $\mathbf{A}^*$, $\bar{\mathbf{A}}$ and $\mathbf{A}^{T}$ are the conjugate transpose, conjugate, and transpose of the matrix $\mathbf{A}$. The $(i,j)$th entry of matrix $\mathbf{A}$ is denoted using $[\mathbf{A}]_{i,j}$. The identity matrix is denoted as $\bI$. 
Further, if $\mathbf{A}$ and $\mathbf{B}$ are two matrices, $\mathbf{A}\circ \mathbf{B}$ is the Khatri-Rao product of $\mathbf{A}$ and $\mathbf{B}$, and $\mathbf{A}\otimes \mathbf{B}$ is their Kronecker product. We use $\mathcal{N}(\mathbf{m},\mathbf{R})$ to denote a circularly symmetric complex Gaussian random vector with mean $\mathbf{m}$ and covariance $\mathbf{R}$. We use $\Exp$ to denote expectation. Discrete time domain signals are represented as $\bx[n]$, with the bold lower case denoting vectors, as before. The frequency domain signals in the $k$th subcarrier are represented using ${\breve{{\boldsymbol{x}}}}[k]$.

\section{System and Channel Models}
In this section, we present the SC-FDE hybrid architecture based system model, followed by a description of the adopted wideband mmWave channel model. The time domain channel estimation algorithm proposed in Section~\ref{sec:sparse_form_time} operates on this kind of SCE-FDE hybrid system, while the frequency domain approach described in Section~\ref{sec:sparse_form_fd} can be applied to OFDM-based hybrid MIMO systems as that in \cite{AlkHea:Frequency-Selective-Hybrid:16}. 

\iftoggle{2column}{
\begin{figure*}
	\centering
	\includegraphics[width=5.5in]{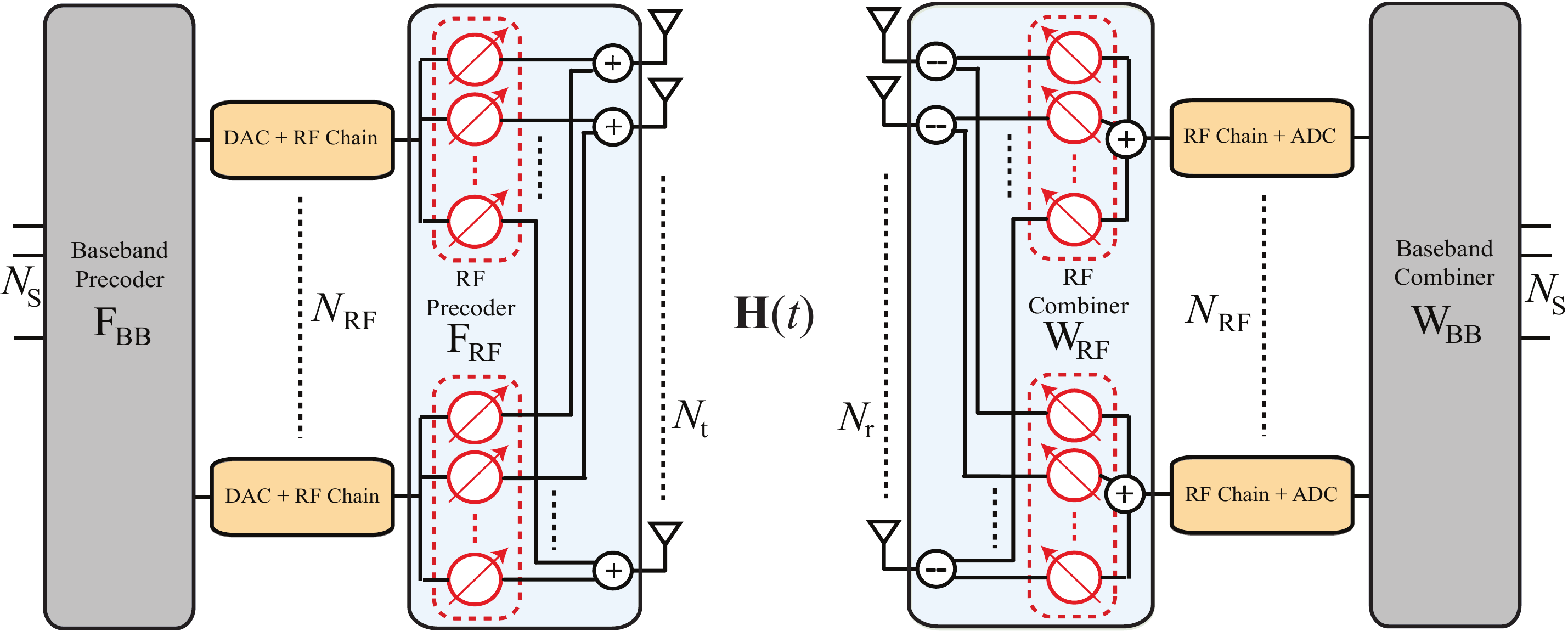} 
	\caption{Figure illustrating the transmitter and receiver structure assumed for the hybrid precoding and combining in the paper. The RF precoder and the combiner are assumed to be implemented using a network of fully connected phase shifters.}     
	\label{fig:hybridModel}        
\end{figure*}
}{
\begin{figure}
	\centering
	\includegraphics[width=5.5in]{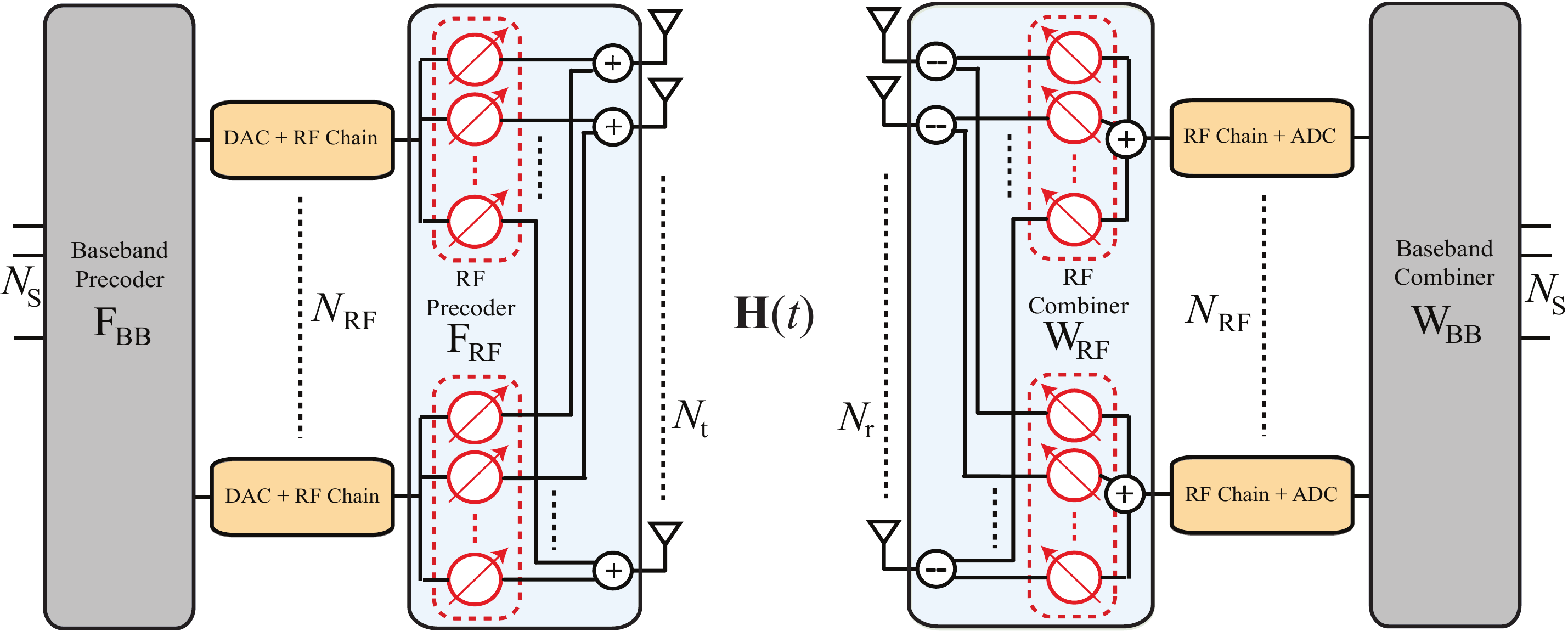} 
	\caption{Figure illustrating the transmitter and receiver structure assumed for the hybrid precoding and combining in the paper. The RF precoder and the combiner are assumed to be implemented using a network of fully connected phase shifters.}     
	\label{fig:hybridModel}        
\end{figure}
}
\subsection{System Model}
Consider a single-user mmWave MIMO system with a transmitter having $\Nt$ antennas and a receiver with $\Nr$ antennas. Both the transmitter and the receiver are assumed to have $\NRF$ RF chains as shown in Fig.~\ref{fig:hybridModel}. The hybrid precoder and combiner used in the frequency selective mmWave system is generally of the form ${\bF}^{\fd}[k] = \FRF{\bF}^{\fd}_{\mathrm{BB}}[k] \in {\bbC}^{\Nt\times\Ns}$ and ${\bW}^{\fd}[k]=\WRF{\bW}^{\fd}_{\mathrm{BB}}[k] \in {\bbC}^{\Nr\times\Ns}$, respectively for the $k$th subcarrier \cite{AlkHea:Frequency-Selective-Hybrid:16}.  In this paper, we focus on the channel estimation having the training precoders/combiners done in the time domain, so we will use $\bF$ and $\bW$ (without $k$) to denote the time domain training precoders/combiners. Accordingly, the transmitter uses a hybrid precoder $\bF = \FRF\FBB \in {\bbC}^{\Nt\times\Ns}$, $\Ns$ being the number of data streams that can be transmitted. Denoting the symbol vector at instance $n$ as ${\bs}[n] \in {\bbC}^{\Ns\times 1}$, satisfying ${\Exp}[{\bs}[n] {\bs}[n]^*] = \frac{1}{\Ns}{\bI}$, the signal transmitted at discrete-time $n$ is $\tilde{\bs}[n] = {\bF}{\bs}[n]$.

The $\Nr \times \Nt$ channel matrix between the transmitter and the receiver is assumed to be frequency selective, having a delay tap length $\Nc$ and is denoted as $\bH_d,~d=1,~2,~...,~\Nc-1$. With 
${\bv}[n] \sim \cN \left(0,\sigma^2 \bI \right)$ denoting the additive noise vector, the received signal can be written as 
\begin{align}
{\br}[n] = \sqrt{\rho}\sum_{d=0}^{\Nc-1}\bH_d {\bF}{\bs}[{n-d}] + {\bv}[n].
\end{align} The noise sample variance $\sigma^2 = \No B$, where $B$ is the wideband system bandwidth, so that the received signal $\SNR = {\rho}/{\sigma^2}$. The receiver applies a hybrid combiner $\bW=\WRF\WBB \in {\bbC}^{\Nr\times\Ns}$, so that the post combining signal at the receiver is 
\begin{align}
{\by}[n] = \sqrt{\rho}\sum_{d=0}^{\Nc-1}{\bW}^*\bH_d {\bF}{\bs}[{n-d}] + {\bW}^*{\bv}[n].
\end{align}

There are several RF precoder and combiner architectures that can be implemented \cite{Mendez-Rial2016}. In this paper, we assume a fully connected phase shifting network \cite{Mendez-Rial2016}. We also consider the constraint so that only quantized angles in 
\begin{align}
{\cA} = \left\lbrace0,~\frac{2\pi}{2^{\NQ}},~\cdots,~\frac{\left(2^{\NQ}-1\right)2\pi}{2^{\NQ}}\right\rbrace \label{eqn:quant_angles}
\end{align} can be realized in the phase shifters. Here $\NQ$ is the number of angle quantization bits. This implies $[{\bF}]_{i,j} = \frac{1}{\sqrt{\Nt}}e^{\jj\varphi_{i,j}}$ and $[{\bW}]_{i,j} = \frac{1}{\sqrt{\Nr}}e^{\jj\omega_{i,j}}$, with $\varphi_{i,j},~\omega_{i,j} \in \cA$.

\subsection{Channel Model}
Consider a geometric channel model \cite{schniter_sparseway:2014, Alkhateeb2014} for the frequency selective mmWave channel consisting of $\Np$ paths. The $d$th delay tap of the channel can be expressed as
\begin{align}
\bH_d = \sum_{\ell = 1}^{\Np}\alpha_{\ell}\prc(d\Ts-\tau_{\ell})\ar(\phi_{\ell})\at^*(\theta_{\ell}), \label{eqn:channel_model}
\end{align}
where $\prc(\tau)$ denotes the raised cosine pulse shaping filter response evaluated at $\tau$, $\alpha_{\ell} \in {\bbC}$ is the complex gain of the $\ell$th channel path, $\tau_{\ell} \in {\bbR}$ is the delay of the $\ell$th path, $\phi_{\ell} \in [0, 2\pi)$ and $\theta_{\ell} \in [0, 2\pi)$ are the angles of arrival and departure, respectively of the $\ell$th path, and $\ar(\phi_{\ell}) \in {\bbC}^{\Nr\times1}$ and $\at(\theta_{\ell}) \in {\bbC}^{\Nt\times1}$ denote the antenna array response vectors of the receiver and transmitter, respectively. 

The transmitter and the receiver are assumed to know the array response vectors. The proposed estimation algorithm applies to any arbitrary antenna array configuration. 
The channel model in \eqref{eqn:channel_model} can be written compactly as 
\begin{align}
\bH_d = \AR \mathbf{\Delta}_d\AT^*, \label{eqn:channel_compact}
\end{align} where $\mathbf{\Delta}_d \in {\bbC}^{\Np\times \Np}$ is diagonal with non-zero entries $\alpha_{\ell}\prc(d\Ts-\tau_{\ell})$, and $\AR \in {\bbC}^{\Nr\times \Np}$ and $\AT \in {\bbC}^{\Nt\times \Np}$ contain the columns $\ar(\phi_{\ell})$ and $\at(\theta_{\ell})$, respectively. Under this notation, vectorizing the channel matrix in \eqref{eqn:channel_compact} gives
\begin{align}
\vect(\bH_d)=\left({\ATbar} \circ \AR\right)\begin{bmatrix}
\alpha_{1}\prc(d\Ts-\tau_{1})\\
\alpha_{2}\prc(d\Ts-\tau_{2})\\
\vdots \\
\alpha_{\Np}\prc(d\Ts-\tau_{\Np})
\end{bmatrix}. \label{eqn:vecHd_expressn}
\end{align} Note that the $\ell$th column of ${\ATbar} \circ \AR$ is of the form ${\atbar}(\theta_{\ell})\otimes\ar(\phi_{\ell})$. We define the vectorized channel
\begin{align}
\hc = \begin{bmatrix}
\vect(\bH_0)\\
\vect(\bH_1)\\
\vdots \\
\vect(\bH_{\Nc-1})
\end{bmatrix}, \label{eqn:vec_channel}
\end{align} which is the unknown signal that is estimated using the channel estimation algorithms proposed in the paper. We assume that the average channel power $\bbE\left[\Vert \hc \Vert^2 _2\right] = \Nr\Nt$ to facilitate comparison of the various channel estimation approaches proposed next.

\section{Time-domain Channel Estimation via Compressed Sensing}
\label{sec:sparse_form_time}
In this section, we present our proposed time-domain explicit channel estimation algorithm that leverages sparsity in the wideband mmWave channel. The hardware constraints on the training frame structure and the precoding-combining beam patterns are also explained.

\subsection{Sparse Formulation in the Time Domain}
For the sparse formulation of the proposed time domain approach, consider block transmission of training frames, with a zero prefix (ZP) appended to each frame \cite{WanMaGia:OFDM-or-SC-block_tx:04, Ghosh2014}. The frame length is assumed to be $N$ and the ZP length is set to $\Nc-1$, with $N > \Nc$, the number of discrete time MIMO channel taps. A hybrid architecture is assumed at the transmitter and the receiver as shown in Fig. \ref{fig:txrx_chain_tdCSModel}. The use of block transmission with $\Nc-1$ zero padding is important here, since it allows reconfiguring the RF circuits from one frame to the other and avoids loss of training data during this reconfiguration. This also avoids inter frame interference. Also note that for symboling rate of 1760 MHZ (the chip rate used in IEEE 802.11ad preamble), it is impractical to use different precoders and combiners for different symbols. It is more feasible, however, to reconfigure the RF circuitry for different frames with $N \sim 16-512$ symbols.

To formulate the sparse recovery problem, we assume that $\NRF$ is the number of RF chains used at the transceivers. For the $m$th training frame, the transmitter uses an RF precoder $\bF_m \in \bbC^{\Nt \times \NRF}$, that can be realized using quantized angles at the analog phase shifters.
\iftoggle{2column}{
\begin{figure*}
\centering
\includegraphics[width=2\columnwidth]{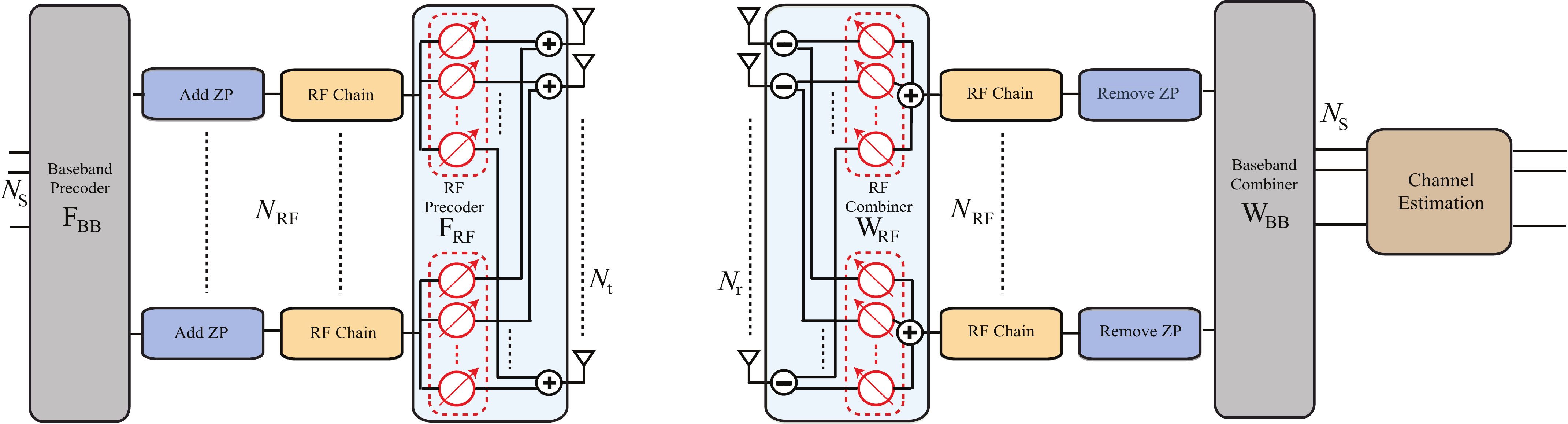} 
\caption{Figure illustrating the transceiver chains and the frame structure assumed for the time-domain channel estimation of the frequency selective mmWave system with $\Nc$ channel taps. Zero padding (ZP) of length at least $\Nc-1$ is prefixed to the training symbols of length $N$ for RF chain reconfiguration across frames.}     
\label{fig:txrx_chain_tdCSModel}        
\end{figure*}
}{
\begin{figure}
\centering
\includegraphics[width=\columnwidth]{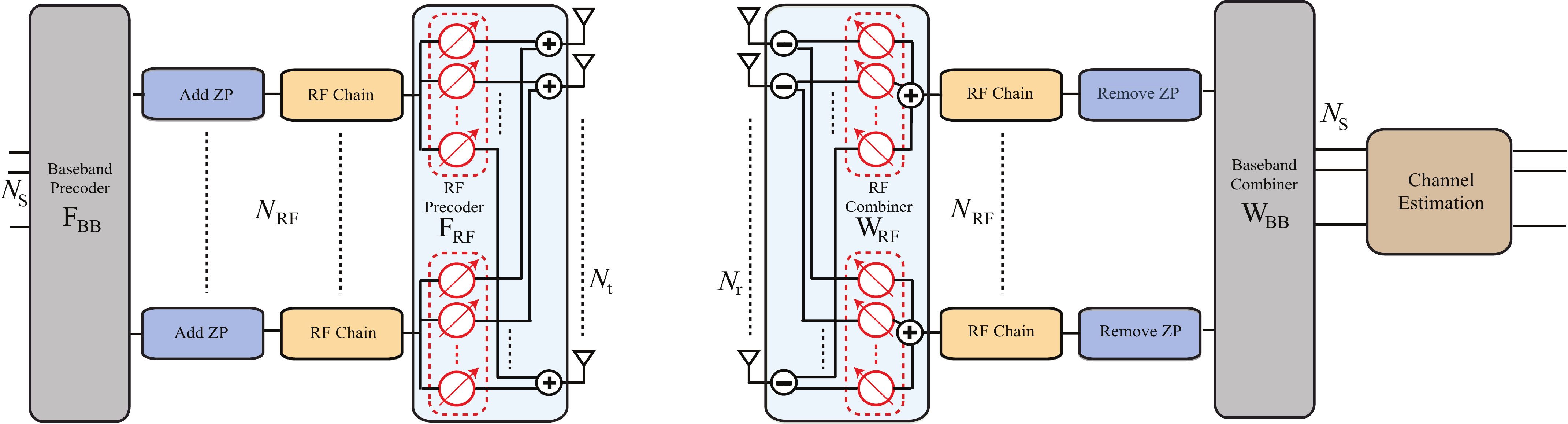} 
\caption{Figure illustrating the transceiver chains and the frame structure assumed for the time-domain channel estimation of the frequency selective mmWave system with $\Nc$ channel taps. Zero padding (ZP) of length at least $\Nc-1$ is prefixed to the training symbols of length $N$ for RF chain reconfiguration across frames.}     
\label{fig:txrx_chain_tdCSModel}        
\end{figure}
} Then, the $n$th symbol of the $m$th received frame is
\begin{align}
{\br}_{m}[n] = \sum_{d=0}^{\Nc-1}\bH_d {\bF}_m {\bs}_{m}[{n-d}] + {\bv}_{m}[n],
\end{align} where ${\bs}_{m}[n] \in \bbC^{\NRF \times 1}$ is the $n$th training data symbol of the $m$th training frame
\begin{align}
{\bs}_{m} = \left[\right.
\underbrace{0~\cdots~0}_{\Nc-1}~\sone~\cdots~\sN \left.
\right].
\end{align}
At the receiver, an RF combiner $\bW_m \in \bbC^{\Nr \times \NRF}$ realized using quantized angles at the analog phase shifters is used during the $m$th training phase. The post combining signal is
\iftoggle{2column}{
\begin{align}
\nonumber
\begin{bmatrix}
\by^{\Trans}_m[1] \\ 
\by^{\Trans}_m[2] \\ 
\vdots \\ 
\by^{\Trans}_m[N]
\end{bmatrix}^\Trans\hspace{-0.13in} = {\bW}_m^*\hspace{-0.04in}\begin{bmatrix}
\bH_0 \hspace{-0.04in}&\hspace{-0.04in}  \cdots \hspace{-0.04in}&\hspace{-0.04in} \bH_{\Nc-1}
\end{bmatrix}\hspace{-0.03in}&\left(\bI_{\Nc}\otimes \bF_m\right){\bS}^{\Trans}_m \\&\hspace{-.25in}+ {\bee}^{\Trans}_m \in \bbC^{1 \times N \NRF},\hspace{-0.05in} \label{eqn:pc_rxsignal}
\end{align}
}
{
\begin{align}
\begin{bmatrix}
\by^{\Trans}_m[1] \\ 
\by^{\Trans}_m[2] \\ 
\vdots \\ 
\by^{\Trans}_m[N]
\end{bmatrix}^\Trans\hspace{-0.13in} = {\bW}_m^*\hspace{-0.04in}\begin{bmatrix}
\bH_0 \hspace{-0.04in}&\hspace{-0.04in}  \cdots \hspace{-0.04in}&\hspace{-0.04in} \bH_{\Nc-1}
\end{bmatrix}\hspace{-0.03in}\left(\bI_{\Nc}\otimes \bF_m\right){\bS}^{\Trans}_m + {\bee}^{\Trans}_m \in \bbC^{1 \times N \NRF},\hspace{-0.05in} \label{eqn:pc_rxsignal}
\end{align}
}
where
\begin{align}
&~\SmBlc = \begin{bmatrix}
\bs^{\Trans}_m[1] & 0  & \cdots &0 \\ 
\bs^{\Trans}_m[2] & \bs^{\Trans}_m[1] & \cdots & .\\ 
\vdots & \vdots & \ddots &\vdots \\ 
\bs^{\Trans}_m[N] & \cdots & \cdots & \bs^{\Trans}_m[{N-\Nc+1}]
\end{bmatrix}, \label{eqn:block_Sm}
\end{align} is of dimension $N \times \Nc\NRF$, and 
\begin{align}
\bbE\left[\bee_{m}\bee^*_m\right] = \sigma^2\bI_{N}\otimes \bW_m^*\bW_m.
\end{align}
Using the matrix equality $\vect\left(\bA\bB\bC\right) = \left(\bC^{\Trans} \otimes \bA \right)\vect\left(\bB\right)$ and the notation for the vectorized channel in \eqref{eqn:vec_channel}, vectorizing \eqref{eqn:pc_rxsignal} gives
\begin{align} {\by}_m =
\begin{bmatrix}
\by_m[1] \\ 
\by_m[2] \\ 
\vdots \\ 
\by_m[N]
\end{bmatrix} =  \underset{{\bf{\Phi}}^{(m)}_{\td}}{\underbrace{\SmBlc \left(\bI_{\Nc}\otimes \bF^{\Trans}_m\right) \otimes {\bW}_m^*}}\hc + {\bee}_{m}. \label{eqn:vec_rxsignal}
\end{align} Using the form in \eqref{eqn:vecHd_expressn} and denoting $\gamma_{\ell,d} = \alpha_{\ell}\prc(d\Ts-\tau_{\ell})$, \eqref{eqn:vec_rxsignal} can be expressed as
\begin{align}
{\by}_{m}= {\bf{\Phi}}^{(m)}_{\td}\left({\bI}_{\Nc}\otimes{\ATbar} \circ \AR \right)\begin{bmatrix}
\gamma_{1,0}\\
\vdots \\
\gamma_{\Np,0}\\
\vdots \\
\gamma_{1,(\Nc-1)}\\
\vdots \\
\gamma_{\Np,(\Nc-1)}
\end{bmatrix}  + {\bee}_{m} \label{eqn:vec_rxsignal_expanded}.
\end{align} In \eqref{eqn:vec_rxsignal_expanded}, the matrices $\AT$ and $\AR$ and the complex gains $\{\alpha_i\}$ and delays $\{\tau_i\}$ contained within $\gamma_{\ell,d}$ are all unknowns that need to be estimated to get the explicit multi tap MIMO channel. Accordingly, we first recover the AoAs / AoDs by estimating the columns of ${\ATbar} \circ \AR$ via sparse recovery.

To formulate the compressed sensing problem in the time domain, we first exploit the sparse nature of the channel in the angular domain. Accordingly, we define the matrices $\Atx$ and $\Arx$ used for sparse recovery, that can be computed apriori at the receiver. The $\Nt \times \Gt$ matrix $\Atx$ consists of columns $\at(\tilde{\theta_x})$, with $\tilde{\theta_x}$ drawn from a quantized angle grid of size $\Gt$, and the $\Nr \times \Gr$ matrix $\Arx$ consists of columns $\ar(\tilde{\phi_x})$, with $\tilde{\phi_x}$ drawn from a quantized angle grid of size $\Gr$. Neglecting the grid quantization error, we can then express \eqref{eqn:vec_rxsignal} as
\begin{align}
{\mathbf{y}}_{m}\hspace{-0.02in}=\hspace{-0.02in} {\bf{\Phi}}^{(m)}_{\td}\left({\bI}_{\Nc}\hspace{-0.02in}\otimes\hspace{-0.02in} {\Atxbar} \hspace{-0.02in}\otimes\hspace{-0.02in} \Arx \hspace{-0.02in}\right)\hat{\bx}_{\td} + {\bee}_{m}. \label{eqn:sparse_form}
\end{align} Note that the actual frequency selective mmWave channel as seen by the RF lens has angles of arrival and departure drawn from $\left[0,2\pi \right)$. The quantization used for constructing the dictionary, when fine enough, can ensure that the dominant AoAs and AoDs are captured as columns of ${\Atxbar} \otimes \Arx$. The error incurred due to the angle grid quantization is investigated in \secref{sec:sim_results}, where we assume offgrid values for the AoA/AoD in the simulations. With this, the signal $\hat{\bx}_{\td}$ consisting of the time domain channel gains and pulse shaping filter response is more sparse than the unknown vector in \eqref{eqn:vec_rxsignal_expanded}, and is of size $\Nc\Gr\Gt \times 1$.

Next, the band-limited nature of the sampled pulse shaping filter is used to operate with an unknown channel vector with a lower sparsity level. For that, we look at the sampled version of the pulse-shaping filter ${\bp}_d$ having entries $p_d(n) = p_{\mathrm{rc}}\left( (d-n)\Ts\right) $, for $d=1,2,~\cdots,~\Nc$ and $n=1,2,~\cdots,~\Gc$. Then, neglecting the quantization error due to sampling in the delay domain, we can write \eqref{eqn:sparse_form} as
\begin{align}
&{\by}_{m}\hspace{-0.02in}={\bf{\Phi}}^{(m)}_{\td}\left({\bI}_{\Nc}\otimes {\Atxbar} \otimes \Arx \right){\bf{\Gamma}}{\bx}_{\td} + {\bee}_{m}, \label{eqn:td_complete_sparse} \\
&\text{where}~~~~~~~~~~~~~~~~
{\bf{\Gamma}} = \begin{bmatrix}
{\bI}_{\Gr\Gt}\otimes{\bp}^{\Trans}_1\\
{\bI}_{\Gr\Gt}\otimes{\bp}^{\Trans}_2\\
\vdots \\
{\bI}_{\Gr\Gt}\otimes{\bp}^{\Trans}_{\Nc}
\end{bmatrix}, \label{eqn:td_Gamma_def}
\end{align} and ${\bx}_{\td} \in \bbC^{\Gc\Gr\Gt \times 1}$ is the $\Np$-sparse vector  containing the time domain complex channel gains.

Stacking $M$ such measurements obtained from sending $M$ training frames and using a different RF precoder and combiner for each frame, we have
\be
&{\by}_{\td} = \,\mathbf{\Phi}_{\td}{\bf{\Psi}}_{\td}\bx_{\td} + \bee, \label{eqn:sparseprblm} \\
&\text{where}~
\by_{\td} = \begin{bmatrix}
\by_1\\
\by_2\\
\vdots \\
\by_M
\end{bmatrix} \in {\bbC}^{NM\NRF \times 1} \label{eqn:measurementstack}
\ee is the measured time domain signal, 
\begin{align}\bbE\left[\bee\bee^*\right] = \sigma^2 \diag\left(\bI_{N}\otimes \bW_1^*\bW_1, \cdots, \bI_{N}\otimes \bW_M^*\bW_M \right), \label{eqn:noise_cov_td}
\end{align}
\begin{align}
\mathbf{\Phi}_{\td} = \begin{bmatrix}
\SBlc \left(\bI_{\Nc}\otimes \bF^{\Trans}_1\right)\otimes {\bW}_1^*\\
{\mathbf{S}}_{2}\left(\bI_{\Nc}\otimes \bF^{\Trans}_2\right)\otimes {\bW}_2^*\\
\hspace{-0.07in}\vdots\\
\SMBlc \left(\bI_{\Nc}\otimes \bF^{\Trans}_m\right)\otimes {\bW}_m^*
\end{bmatrix} \in {\bbC}^{NM\NRF\times \Nc\Nr\Nt} \label{eqn:sensemat}
\end{align}
 is the time domain measurement matrix, and
\begin{align}
\mathbf{\Psi}_{\td} &= \left({\bI}_{\Nc}\otimes\ {\Atxbar}\otimes \Arx \right){\bf{\Gamma}} \nonumber\\
&= \begin{bmatrix}
\left( {\Atxbar}\otimes \Arx \right)\otimes{\bp}^{\Trans}_1\\
\left( {\Atxbar}\otimes \Arx \right)\otimes{\bp}^{\Trans}_2\\
\hspace{-0.07in}\vdots\\
\left( {\Atxbar}\otimes \Arx \right)\otimes{\bp}^{\Trans}_{\Nc}
\end{bmatrix} \in {\bbC}^{\Nc\Nr\Nt \times \Gc\Gr\Gt} \label{eqn:td_dictionary}
\end{align} is the dictionary in the time domain. The beamforming and combining vectors ${\bF}_m,~{\bW}_m$, $m=1,~2,~\cdots,~M$ used for training have the phase angles chosen uniformly at random from the set $\cA$ in \eqref{eqn:quant_angles}.

\subsection{AoA/AoD and Channel Gain Estimation in the Time Domain}
\label{ssec:Sparse_recovery_td}
With the sparse formulation of the mmWave channel estimation problem in \eqref{eqn:sparseprblm}, compressed sensing tools can be first used to estimate the AoA and AoD. The support of $\bx_{\td}$ corresponds to a particular AoA, AoD and path delay, and hence estimating the support of $\bx_{\td}$ amounts to estimating a channel path, and the corresponding non-zero value corresponds to the path gain. Note that we can increase or decrease the angle quantization grid sizes $\Gr$ and $\Gt$, and the delay domain quantization grid size $\Gc$, used for constructing the time domain dictionary to minimize the quantization error. As the sensing matrix is known at the receiver, sparse recovery algorithms can be used to estimate the AoA and AoD. 

To estimate the support of the sparse vector $\bx_{\td}$, we solve the optimization problem
\begin{align}
\underset{\bx_{\td}}{\min}~ \Vert \bx_{\td} \Vert_1 \quad\mathrm{such~that}\quad \Vert \by_{\td}  - {\bf{\Phi}}_{\td} {\bf{\Psi}}_{\td}\bx_{\td} \Vert_2 \leq \epsilon. \label{eqn:cs_fund_cvx}
\end{align} We consider Orthogonal Matching Pursuit (OMP) for solving \eqref{eqn:cs_fund_cvx}, as used previously in \cite{Alkhateeb2015,TauHlaEiwRau:Compressive-estimation-of-doubly-selective:10}.
There are several stopping criteria for OMP that can be used to solve \eqref{eqn:cs_fund_cvx}. When the sparsity level $\Np$ is known apriori, reaching that level could be used to stop the algorithm. When such information cannot be guessed before hand (which itself is an estimation problem), the residual error falling below a certain threshold is often used to terminate the recursive OMP algorithm. Accordingly, in the presence of noise, a suitable choice for the threshold $\epsilon$ is the noise variance. Hence, we assume the noise power as the stopping threshold, i.e.,
$\epsilon = \bbE [\bee^*\bee]$.

Following the support estimation via sparse recovery, the channel gains can be estimated. While there are many ways to estimate the gains, even directly from OMP, we only give the details for one approach next -- using least squares. The various methods are based on plugging in the columns of the dictionary matrices corresponding to the estimated AoA and AoD. That is, let $\cS^{\td}_{\sfA}$ and $\cS^{\td}_{\sfD}$, respectively be the estimated AoA and AoD using sparse recovery in the proposed time domain formulation. Then, using \eqref{eqn:sparse_form} and stacking the $M$ measurements, we have
\begin{align}
{\by}_{\td}= \underset{\bf{\Omega}_{\td}}{\underbrace{\mathbf{\Phi}_{\td} \left({\bI}_{\Nc}\otimes\left[ {\Atxbar}\right]_{:,\cS^{\td}_{\sfD}} \otimes\left[ \Arx \right]_{:,\cS^{\td}_{\sfA}}\right)}}\hat{\bx}_{\td} + {\mathbf{e}}, \label{eqn:sparse_reduced}
\end{align} so that the channel coefficients via least squares is
\begin{align}
\hat{\bx}^{\mathrm{LS}}_{\td} = \left(\bf{\Omega}^*_{\td}\bf{\Omega}_{\td}\right)^{-1}\bf{\Omega}^*_{\td}{\by}_{\td}. 
\end{align}

\section{Frequency-domain Channel Estimation via Compressed Sensing}
\label{sec:sparse_form_fd}
In this section, we explain how the compressed sensing problem can be formulated in the frequency domain. The additional modifications needed in the system model, and the corresponding advantages and disadvantages are also explained in this section. 

Using the geometric channel model in \eqref{eqn:channel_model}, the complex channel matrix in the frequency domain can be written as
\iftoggle{2column}{
\begin{align}
\nonumber&{\fbH}[k] = \sum_{d=0}^{\Nc-1} {\bH}_d e^{-\jj\frac{2\pi k d}{K}} \\
&=\sum_{\ell=1}^\Np \alpha_{\ell}\ar(\phi_{\ell})\at^*(\theta_{\ell}) \sum_{d=0}^{\Nc-1}\prc(d\Ts-\tau_{\ell})e^{-\jj\frac{2\pi k d}{K}}.
\end{align}
}{
\begin{align}
\nonumber{\fbH}[k] &= \sum_{d=0}^{\Nc-1} {\bH}_d e^{-\jj\frac{2\pi k d}{K}} \\
&= \sum_{\ell=1}^\Np \alpha_{\ell}\ar(\phi_{\ell})\at^*(\theta_{\ell})\sum_{d=0}^{\Nc-1}\prc(d\Ts-\tau_{\ell})e^{-\jj\frac{2\pi k d}{K}}.
\end{align}
}
Defining $\beta_{k,\ell} = \sum_{d=0}^{\Nc-1}\prc(d\Ts-\tau_{\ell})e^{-\jj\frac{2\pi k d}{K}}$ a compact expression can be derived
\begin{align}
{\fbH}[k] = \sum_{\ell=1}^\Np \alpha_{\ell}\beta_{k,\ell}\ar(\phi_{\ell})\at^*(\theta_{\ell}). \label{eqn:Hk_fd}
\end{align} Vectorizing \eqref{eqn:Hk_fd} gives the unknown signal that is estimated using the frequency domain estimation algorithm,
\begin{align}
\vect\left({\fbH}[k]\right) = \left({\ATbar} \circ \AR\right)\begin{bmatrix}
\alpha_{1}\beta_{k,1}\\
\alpha_{2}\beta_{k,2}\\
\vdots \\
\alpha_{\Np}\beta_{k,\Np}
\end{bmatrix}. \label{eqn:vecHk_fd_expressn}
\end{align} Note that the vector channel representation of the $k$th subcarrier in \eqref{eqn:vecHk_fd_expressn}, is similar to the time domain vector representation in \eqref{eqn:vecHd_expressn}. The key difference, however, is that, unlike the time domain approach, each of the unknown vectors corresponding to the $K$ subcarriers can be estimated separately, in parallel as explained next.

\subsection{Sparse Formulation in the Frequency Domain}
\label{sec:sparse_form_frequency}
For the sparse formulation in the proposed frequency domain approach, we assume that appropriate signal processing is performed to convert the linear convolution occurring during the frame transmission in the system to a circular convolution in the time domain. That is, with ZP assumed in the system model and the time domain frame structure for the training preamble, the overlapping and sum \cite{CheZhaJay:New-training-seq-ZP_SC-FDE:08} method is used, followed by the $K$-point FFT to formulate the frequency-domain sparse channel estimation problem per subcarrier $k = 1, 2, \cdots, K$. The proposed system model with the hybrid architecture and signal processing components required for the frequency domain channel estimation in an SC-FDE system with ZP is illustrated in Fig. \ref{fig:txrx_chain_fdCSModel}. As in the proposed time domain approach in \secref{sec:sparse_form_time}, prefixing zeros (ZP) to each frame facilitates reconfiguration of the RF precoders and combiners from frame to frame. Alternatively, for an OFDM based system, cyclic prefixing (CP) is performed at the transmitter, which is discarded at the receiver before the FFT operation. The advantage of the proposed frequency domain approach is that different baseband precoders and combiners can be used for different subcarriers \cite{AlkHea:Frequency-Selective-Hybrid:16} in the frequency domain, while the RF processing is frequency flat. The proposed frequency domain approach, therefore, works for both SC-FDE and OFDM systems, where the received signal is processed per subcarrier. We next look into the received signal in the $k$th subcarrier.

With $\bF_m$ denoting the RF precoder used at the transmitter for the transmission of the $m$th training frame/OFDM symbol, and $\bW_m$, the corresponding RF combiner, the post combining signal in the $k$th subcarrier can be written as
\begin{align}
{\breve{{\boldsymbol{y}}}}_{m}[k] &= {\bW^*_m}{\fbH}[k] {\bF}_m \breve{\boldsymbol{s}}_{m}[k] + \breve{\boldsymbol{e}}_{m}[k], \label{eqn:freq_received_y}\\
\text{where} \quad 
\breve{\boldsymbol{s}}_{m}[k] &= \sum_{n=1}^{N} \bs_{m}[n] e^{-\jj\frac{2\pi k n}{K}}
\end{align} is the $k$th coefficient of the $K$-point FFT of $m$the time domain transmit frame. The covariance of the frequency domain noise vector in \eqref{eqn:freq_received_y} is $\bbE\left[\breve{\boldsymbol{e}}_{m}[k]\breve{\boldsymbol{e}}^*_m[k]\right] = \sigma^2\bW^*_m\bW_m$, and $\sigma^2 = \No B.$ The frequency flat RF combiners and precoders are assumed to be realized with a network of phase shifters with phase angles drawn from a finite set, as before.
\iftoggle{2column}{
\begin{figure*}
\centering
\includegraphics[width=2\columnwidth]{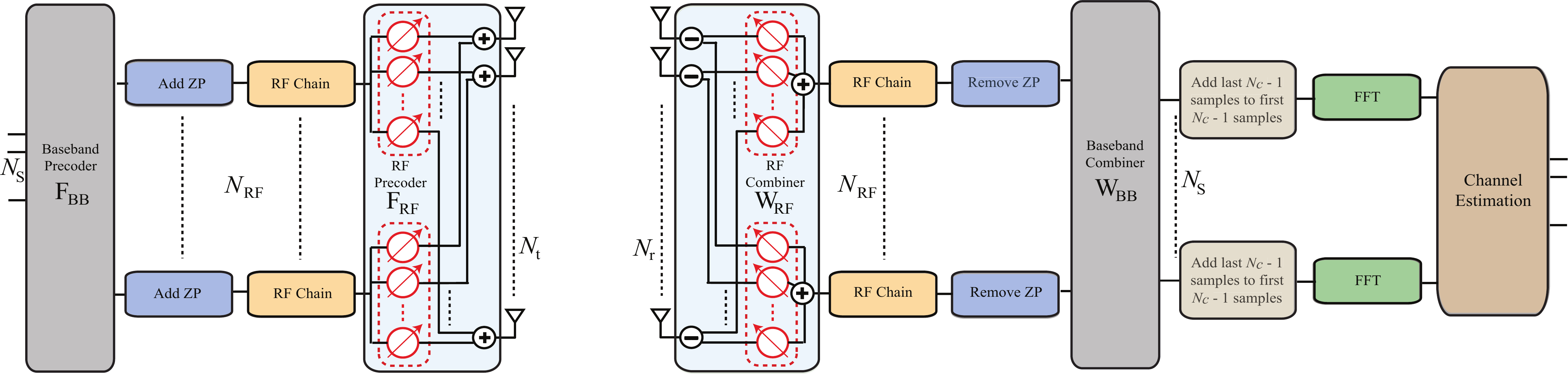} 
\caption{Figure illustrating the transceiver chains and the frame structure assumed for the frequency-domain channel estimation of the frequency selective mmWave system with $\Nc$ channel taps. Zero padding (ZP) of length $\Nc-1$ is prefixed to the training symbols of length $N$ for RF chain reconfiguration across frames. }     
\label{fig:txrx_chain_fdCSModel}        
\end{figure*}
}{
\begin{figure}
\centering
\includegraphics[width=\columnwidth]{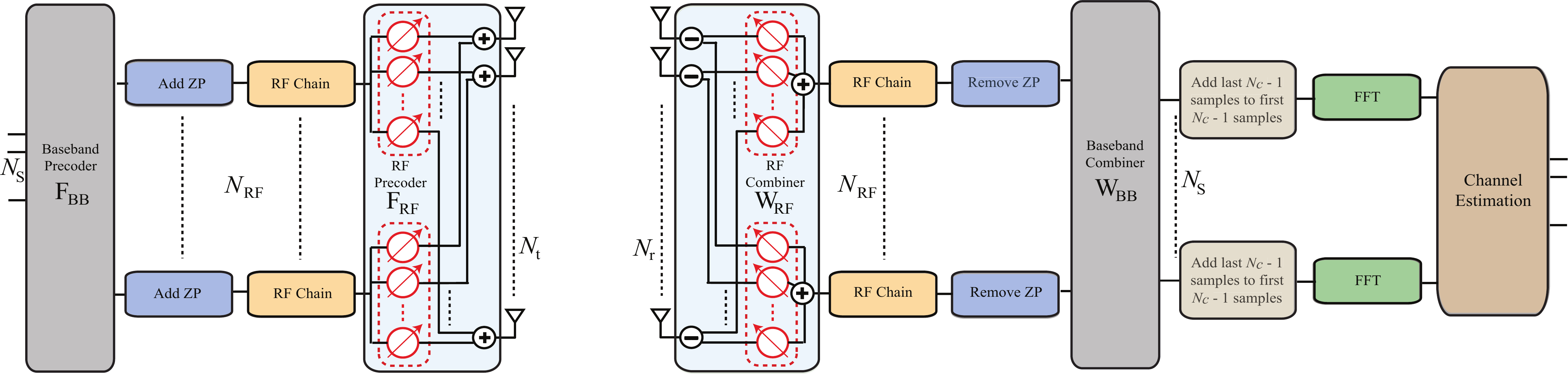} 
\caption{Figure illustrating the transceiver chains and the frame structure assumed for the frequency-domain channel estimation of the frequency selective mmWave system with $\Nc$ channel taps. Zero padding (ZP) of length $\Nc-1$ is prefixed to the training symbols of length $N$ for RF chain reconfiguration across frames. }     
\label{fig:txrx_chain_fdCSModel}        
\end{figure}
}
Vectorizing \eqref{eqn:freq_received_y}, and substituting \eqref{eqn:vecHk_fd_expressn} gives
\begin{align}
\vect\left({\breve{\boldsymbol{y}}}_{m}[k]\right) = \underset{{\bf{\Phi}}^{(m)}_{\fd}[k]}{\underbrace{\left(\breve{\boldsymbol{s}}^{\Trans}_m[k] {\bF}^\Trans_m \otimes {\bW^*_m}\right)}} \vect\left({\fbH}[k]\right) + \breve{\boldsymbol{e}}_{m}[k]. \label{eqn:vecyk_fd}
\end{align}
 Assuming the AoAs and AoDs are drawn from a grid of size $\Gr$ and $\Gt$, respectively, and neglecting the quantization error, we can write \eqref{eqn:vecyk_fd} in terms of the dictionary matrices defined in Section \ref{sec:sparse_form_time} as follows:
\iftoggle{2column}{
\begin{align}
\vect\left({\breve{\boldsymbol{y}}}_{m}[k]\right) = {\bf{\Phi}}^{(m)}_{\fd}[k]\left({\Atxbar} \otimes \Arx \right)\breve{\boldsymbol{x}}[k] + \breve{\boldsymbol{e}}_{m}[k],
\end{align}
}{
\begin{align}
\vect\left({\breve{\boldsymbol{y}}}_{m}[k]\right) = {\bf{\Phi}}^{(m)}_{\fd}[k]\left({\Atxbar} \otimes \Arx \right)\breve{\boldsymbol{x}}[k] + \breve{\boldsymbol{e}}_{m}[k],
\end{align}
} with the signal $\breve{\boldsymbol{x}}[k] \in {\mathbb{C}}^{\Gr\Gt \times 1}$ being $\Np$-sparse. Stacking $M$ such measurements obtained over the course of $M$ training frame transmission, each with a different pair of RF precoder and combiner, we have the following sparse formulation for the $k$th subcarrier
\begin{align}
{\breve{\boldsymbol{y}}}[k] = {\mathbf{\Phi}}_{\fd}[k]{\mathbf{\Psi}}_{\fd}\breve{\boldsymbol{x}}[k] + \breve{\boldsymbol{e}}[k], \label{eqn:sparse_fd_Mstack}
\end{align} in terms of the frequency domain dictionary ${\mathbf{\Psi}}_{\fd} = \left({\Atxbar} \otimes \Arx \right) \in \bbC^{\Nr\Nt \times \Gr\Gt}$ and the measurement matrix in the frequency domain
\begin{align}
{\mathbf{\Phi}}_{\fd}[k] = \begin{bmatrix}
\breve{\boldsymbol{s}}^{\Trans}_1[k]{\bF}^{\Trans}_1 \otimes {\bW^*_1} \\
\breve{\boldsymbol{s}}^{\Trans}_2[k]{\bF}^{\Trans}_2 \otimes {\bW^*_2} \\
\hspace{-0.07in}\vdots\\
\breve{\boldsymbol{s}}^{\Trans}_M[k]{\bF}^{\Trans}_M \otimes {\bW^*_M} 
\end{bmatrix} \in \bbC^{M\NRF \times \Nr\Nt}. 
\end{align} The covariance of the noise  in \eqref{eqn:sparse_fd_Mstack} is
\begin{align}
\bbE\left[\breve{\boldsymbol{e}}[k]\breve{\boldsymbol{e}}^*[k]\right] = \sigma^2\diag\left(\bW^*_1\bW_1, \bW^*_2\bW_2,\cdots,\bW^*_M\bW_M\right). \label{eqn:noise_cov_fd}
\end{align}

\subsection{AoA/AoD and Channel Gain Estimation per Subcarrier}
\label{ssec:Sparse_recovery_fd}
As discussed previously in \secref{ssec:Sparse_recovery_td}, we first estimate the support of ${\breve{\boldsymbol{x}}}[k]$, that corresponds to a particular AoA and AoD, and then proceed to estimate the MIMO channel coefficients of the $k$th subcarrier, which correspond to the non-zero values of ${\breve{\boldsymbol{x}}}[k]$. As with the time domain approach, we solve the following optimization problem
\begin{align}
\underset{{\breve{\boldsymbol{x}}}[k]}{\min}~ \Vert {\breve{\boldsymbol{x}}}[k] \Vert_1 \quad\mathrm{such~that}\quad \Vert {\breve{\boldsymbol{y}}}[k]  - {\bf{\Phi}}_{\fd}[k] {\bf{\Psi}}_{\fd}{\breve{\boldsymbol{x}}}[k] \Vert_2 \leq \epsilon. \label{eqn:cs_fund_cvx_fd}
\end{align} via OMP with the stopping threshold $\epsilon = \bbE [ \breve{\boldsymbol{e}}[k]^* \breve{\boldsymbol{e}}[k]]$,
to estimate the support of the sparse vector ${\breve{\boldsymbol{y}}}[k]$, and hence the dominant angles of arrival and departure. The set of estimated AoAs is denoted as $\cS^{\fd}_{\sfA}$, and the set of AoDs is denoted as $\cS^{\fd}_{\sfD}$. These sets correspond to specific columns of the frequency domain dictionary $\Psi_{\fd}$. Using $\cS^{\fd}_{\sfA}$ and $\cS^{\fd}_{\sfD}$, the channel coefficients, that correspond to the non-zero values of the sparse vector $\breve{\boldsymbol{x}}[k],$ can be derived as follows. From \eqref{eqn:sparse_fd_Mstack}, after the sparse angle recovery
\begin{align}
{\breve{\boldsymbol{y}}}[k] = \underset{\bf{\Omega}_{\fd}}{\underbrace{\mathbf{\Phi}_{\fd}[k]\left(\left[{\Atxbar}\right]_{:,\cS^{\fd}_{\sfD}} \otimes \left[\Arx\right]_{:,\cS^{\fd}_{\sfA}} \right)}}\breve{\boldsymbol{x}}[k] + \breve{\boldsymbol{e}}[k],
\end{align} so that, using least square estimation,
\begin{align}
\breve{\boldsymbol{x}}^{\mathrm{LS}}[k] = \left(\bf{\Omega}^*_{\fd}\bf{\Omega}_{\fd}\right)^{-1}\bf{\Omega}^*_{\fd}{\breve{\boldsymbol{y}}}[k].
\end{align}

Note that using the sparse formulation in \eqref{eqn:sparse_fd_Mstack}, the AoAs/AoDs  and the channel coefficients of the the $k$th subcarrier can be estimated. Repeating the same for all the $K$ subcarriers fully characterizes the frequency selective mmWave channel. While the dimensions of the matrices involved in the frequency-domain compressed sensing problem is smaller in comparison to the time-domain formulation in Section \ref{sec:sparse_form_time}, the channel estimation should be invoked $K$ times to fully recover the channel coefficients. Further, additional pre-processing and FFT operation are required.

\section{Combined Time-Frequency Compressive Channel Estimation}
In this section, we formulate a technique via compressed sensing for explicit channel estimation, jointly in time and frequency. The key idea is to estimate the angles of arrival and departure via compressed sensing in the frequency domain, and then use the estimates to evaluate the channel gains and path delays in the time domain to obtain the entire channel.

The transmitter chain for the proposed combined time-frequency compressive channel estimation approach is the same as in \figref{fig:txrx_chain_tdCSModel} and \figref{fig:txrx_chain_fdCSModel}. The system model for the receiver chain in \figref{fig:txrx_chain_fdCSModel}, can be employed to perform sparse support recovery of the angles in the frequency domain for the proposed estimation approach in this section. Following the compressive support estimation, the pre-computed dictionary matrices in the time domain, and the measurement matrices can be used to estimate the channel coefficients of the frequency selective mmWave MIMO channel, as explained momentarily.

From \eqref{eqn:vecyk_fd} and \eqref{eqn:vecHk_fd_expressn}, we can express the frequency domain received signal in the $k$th subcarrier, in terms of the actual AoAs and AoDs in the vector form as follows
\begin{align}
\vect\left({\breve{\boldsymbol{y}}}_{m}[k]\right) = {\bf{\Phi}}^{(m)}_{\fd}[k]\left({\ATbar} \circ \AR\right)\begin{bmatrix}
\alpha_{1}\beta_{k,1}\\
\alpha_{2}\beta_{k,2}\\
\vdots \\
\alpha_{\Np}\beta_{k,\Np}
\end{bmatrix}  + \breve{\boldsymbol{e}}_{m}[k], \label{eqn:joint_freq_1}
\end{align}
 with the noise covariance $\bbE\left[\breve{\boldsymbol{e}}_{m}[k]\breve{\boldsymbol{e}}^*_{m}[k]\right] = \sigma^2\bW^*_m\bW_m$.

\textit{AoA/AoD Estimation in Frequency Domain and Channel Gain Estimation in Time Domain}: 
Note from \eqref{eqn:joint_freq_1}, the AoA and AoD information in each subcarrier $k$ is the same, and contained in ${\ATbar} \circ \AR$, whose $\ell$th column is of the form ${\atbar}(\theta_{\ell})\otimes\ar(\phi_{\ell})$. Therefore, using sparse recovery in $1 \leq P \leq K$ number of subcarriers parallely, and concatenating the estimated angles, we can get a support set of the AoAs, denoted as $\cS_{\sfA}$ and a set of AoD estimates denoted as $\cS_{\sfD}$. One option to estimate the support is to use OMP as explained in \secref{ssec:Sparse_recovery_fd}, for $P$ subcarriers in parallel. Prior work in \cite{GaoDai2ShiWan:Structured-compressive-sensing-based:16, WanDaiMirWan:Joint-user-activity-data-detection:16,GaoHuDaiWan:Channel-estimation-mmWaveMassiveMIMO:16} have studied methods to estimate a common support set from multiple parallel measurements. Such techniques may also be employed to recover the support set containing the AoA/AoD information in the frequency domain, both here as well as in the proposed approach in \secref{sec:sparse_form_fd}.

Following the support recovery in the frequency domain, to recover the entire channel, we switch to the time domain formulation in \eqref{eqn:sparse_form}, \eqref{eqn:measurementstack}, \eqref{eqn:sensemat} and \eqref{eqn:td_dictionary}, but restrict to the set $\mathcal{S} = \left\lbrace\cS_{\sfA},\cS_{\sfD}\right\rbrace$. Accordingly, we can write the effective time-domain equation, conditioned on the estimated support set $\mathcal{S}$ as
\begin{align}
{\by}_{\td} = \underset{\bf{\Omega}}{\underbrace{\mathbf{\Phi}_{\td}\left[{\bf{\Psi}}_{\td}\right]_{:,\mathcal{S}}}}\bx_{\td} + \bee, \label{eqn:joint_td_fd_conditional}
\end{align} where $\mathbf{\Phi}_{\td}$ is that in \eqref{eqn:sensemat}, the noise covariance of $\bee$ is that in \eqref{eqn:noise_cov_td}, and 
\begin{align}
\left[{\bf{\Psi}}_{\td}\right]_{:,\mathcal{S}} = \begin{bmatrix}
\left[ {\Atxbar}\right]_{:,\cS_{\sfD}}\otimes \left[\Arx \right]_{:,\cS_{\sfA}}\otimes{\bf{\tilde{p}}}^T_1\\
\left[ {\Atxbar}\right]_{:,\cS_{\sfD}}\otimes \left[\Arx \right]_{:,\cS_{\sfA}}\otimes{\bf{\tilde{p}}}^T_2\\
\hspace{-0.07in}\vdots\\
\left[ {\Atxbar}\right]_{:,\cS_{\sfD}}\otimes \left[\Arx \right]_{:,\cS_{\sfA}}\otimes{\bf{\tilde{p}}}^T_{\Nc}
\end{bmatrix} \in {\mathbb{C}}^{\Nc\Nr\Nt \times \Gc\vert\mathcal{S}\vert} 
\end{align} is the dictionary matrix conditioned on the knowledge of the support set. The unknown $\mathbf{\hat{x}}$ in \eqref{eqn:joint_td_fd_conditional}, contains channel coefficients in the time domain, which can now be obtained via least squares or MMSE to recover the entire MIMO channel matrices corresponding to all the delay taps.  That is, from \eqref{eqn:joint_td_fd_conditional}
\begin{align}
\bx^{\mathrm{LS}}_{\td} = \left(\bf{\Omega}^*\bf{\Omega}\right)^{-1}\bf{\Omega}^*{\by}_{\td}.
\end{align}

The advantage of using the combined time-frequency approach for the wideband channel estimation is twofold. First, since the sparse recovery is done in the frequency domain, the sizes of the measurement matrix and the dictionary are $M\NRF \times \Nr\Nt$ and $\Nr\Nt \times \Gr\Gt$, respectively, that are much smaller than the corresponding time domain matrices ${\bf{\Phi}}_{\td}$ and ${\bf{\Psi}}_{\td}$. Secondly, unlike the frequency domain approach, the channel estimates need not be separately evaluated per subcarrier, but only once in the time domain, thus further reducing the computation complexity.

\section{Simulation Results}
\label{sec:sim_results}
In this section, the performance of the three proposed channel estimation algorithm are provided. For the compressed sensing estimation of the angles of arrival and departure, orthogonal matching pursuit is used. The channel gains are then estimated using least squares.

We assume uniform linear array (ULA) with half wavelength antenna element separation for the simulations. For such a ULA,
\begin{align} 
\nonumber \ar(\phi_{\ell})= \frac{1}{\sqrt{\Nr}}\begin{bmatrix}1 &e^{\jj\pi\cos\left(\phi_{\ell}\right)}&\cdots&e^{\jj(\Nr-1)\pi\cos\left(\phi_{\ell}\right)}\end{bmatrix}^T , 
\end{align} and
\begin{align}
\nonumber \at(\theta_{\ell}) =\frac{1}{\sqrt{\Nt}}\begin{bmatrix}1 &e^{\jj\pi\cos\left(\theta_{\ell}\right)}&\cdots&e^{\jj(\Nt-1)\pi\cos\left(\theta_{\ell}\right)}\end{bmatrix}^T . 
\end{align} 
The AoA and AoD quantization used for constructing the transmitter and receiver dictionary matrices are taken from an angle grid of size $\Gr$ and $\Gt$, respectively. This implies that the $\ell$th column of $\Atx$ is $\at(\tilde{\theta_\ell})$, where $\tilde{\theta_\ell} = \frac{(\ell-1)\pi}{\Gt}$ and the $k$th column of $\Arx$ is $\ar(\tilde{\phi_k})$, where $\tilde{\phi_k} = \frac{(k-1)\pi}{\Gr}$ . The angle quantization used in the phase shifters is assumed to have $\NQ$ quantization bits, so that the entries of $\bF_m,~\bW_m$, $m=1,~2,~\cdots,~M$ are drawn from $\cA$, as defined in \eqref{eqn:quant_angles}, with equal probability. The $\Np$  paths of the wideband mmWave channel are assumed to be independently and identically distributed, with delay $\tau_{\ell}$ chosen uniformly at random from $[0, (\Nc-1)\Ts]$, where $\Ts$ is the sampling interval and $\Nc$ is the number of delay taps of the channel. The angles of arrival and departure for each of the channel paths are assumed to be distributed independently and uniformly in $[0,\pi]$. The raised cosine pulse shaping signal is assumed to have a roll-off factor of $0.8$.

\iftoggle{2column}{
\begin{figure}
	\centering
	\includegraphics[width=3.25in,height=2.6in]{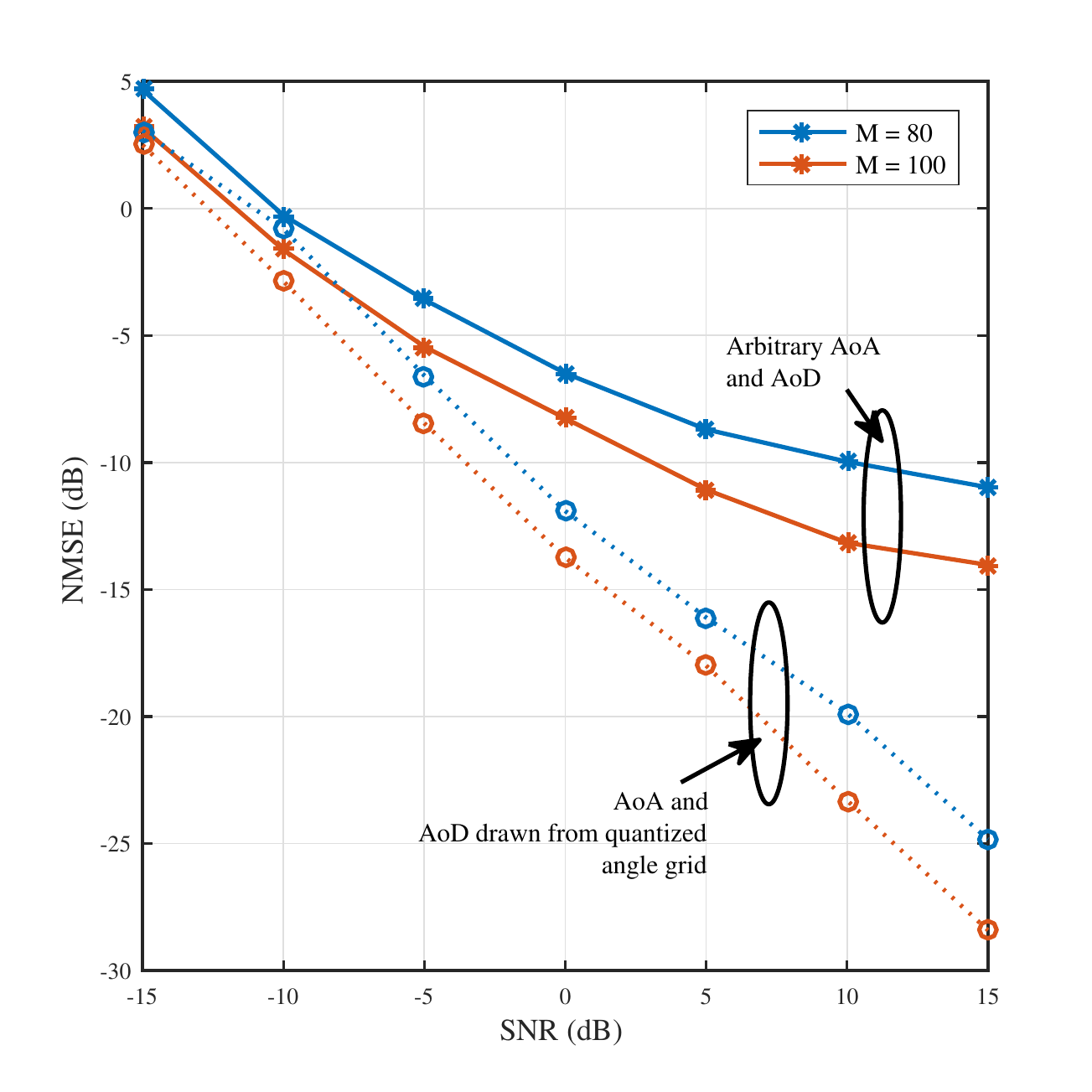} 
	\caption{Average $\ttNMSE$ as a function of SNR for different training length $M$ when $N_s = 1$ and $\NRF = 1$ using the proposed time-domain channel estimation technique. We assume $N=16$ symbols per frame for a frequency selective channel of 4 taps. Using the proposed approach, training length of $80-100$ is sufficient to ensure very low estimation error, processing completely in the time-domain.}
	\label{fig:NMSEwithMeas}        
\end{figure}
}{
\begin{figure}
	\centering
	\includegraphics[width=5in,height=4in]{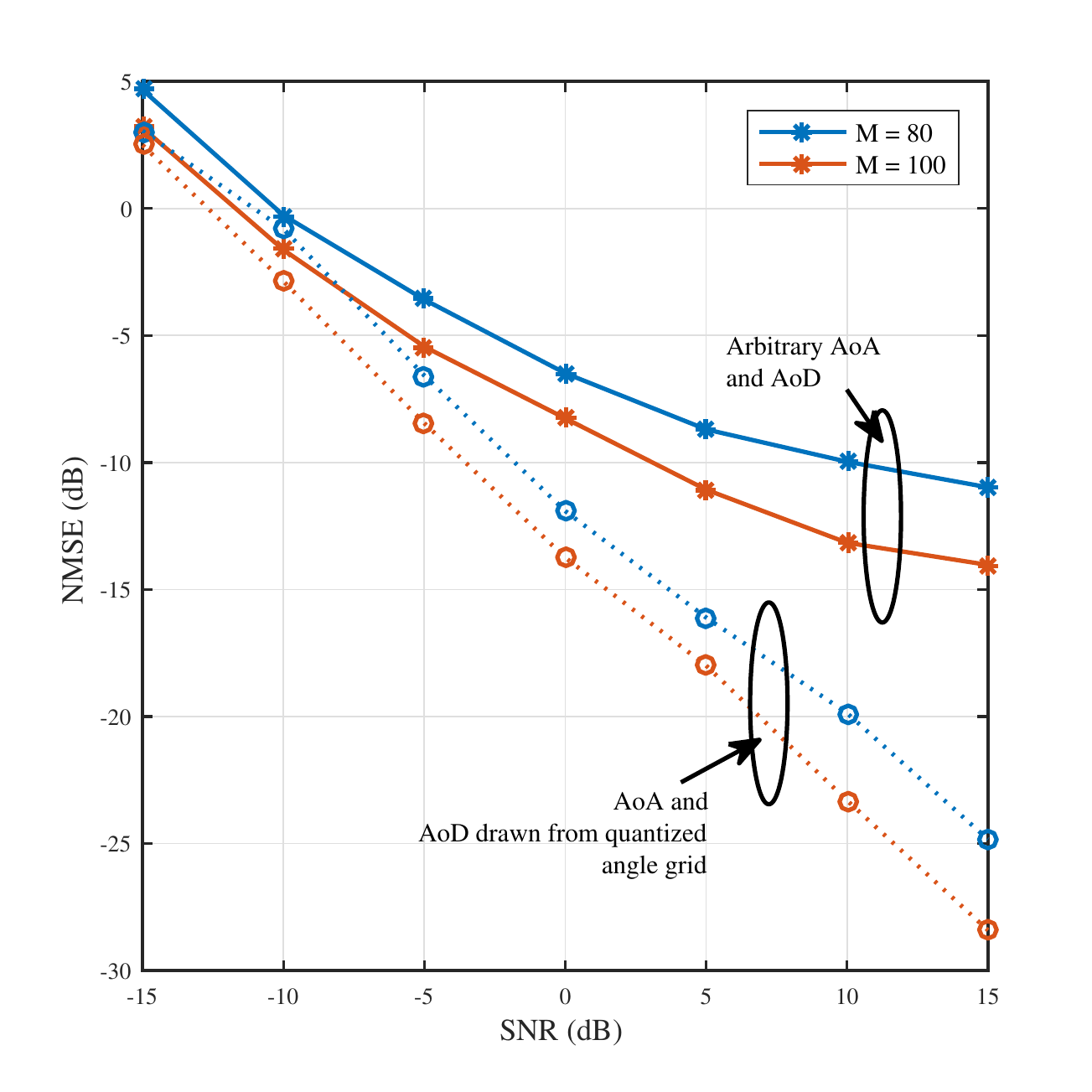} 
	\caption{Average $\ttNMSE$ as a function of SNR for different training length $M$ when $N_s = 1$ and $\NRF = 1$ using the proposed time-domain channel estimation technique. We assume $N=16$ symbols per frame for a frequency selective channel of 4 taps. Using the proposed approach, training length of $80-100$ is sufficient to ensure very low estimation error, processing completely in the time-domain.}
	\label{fig:NMSEwithMeas}        
\end{figure}
}
Let $\hchat \in \bbC^{\Nc\Nr\Nt \times 1}$ denote the estimated channel vector. We use the following metrics to compare the performance of our proposed channel estimation algorithms:
\begin{itemize}
\item[1)] the normalized mean squared error ($\ttNMSE$) of the channel estimates defined as
\begin{align}
\ttNMSE = \frac{\Vert \hc - \hchat \Vert^2_2}{\Vert\hc\Vert^2_2} = \frac{\sum_{d=0}^{\Nc-1} ||\bH_d - \hat{\bH_d} ||^2_F}{\sum_{d=0}^{\Nc-1} ||\bH_d||^2_F}.
\end{align}
\item[2)] the ergodic spectral efficiency as defined in \cite{schniter_sparseway:2014}.
\end{itemize}

\iftoggle{2column}{
\begin{figure}
	\centering
	\includegraphics[width=3.25in,height=2.6in]{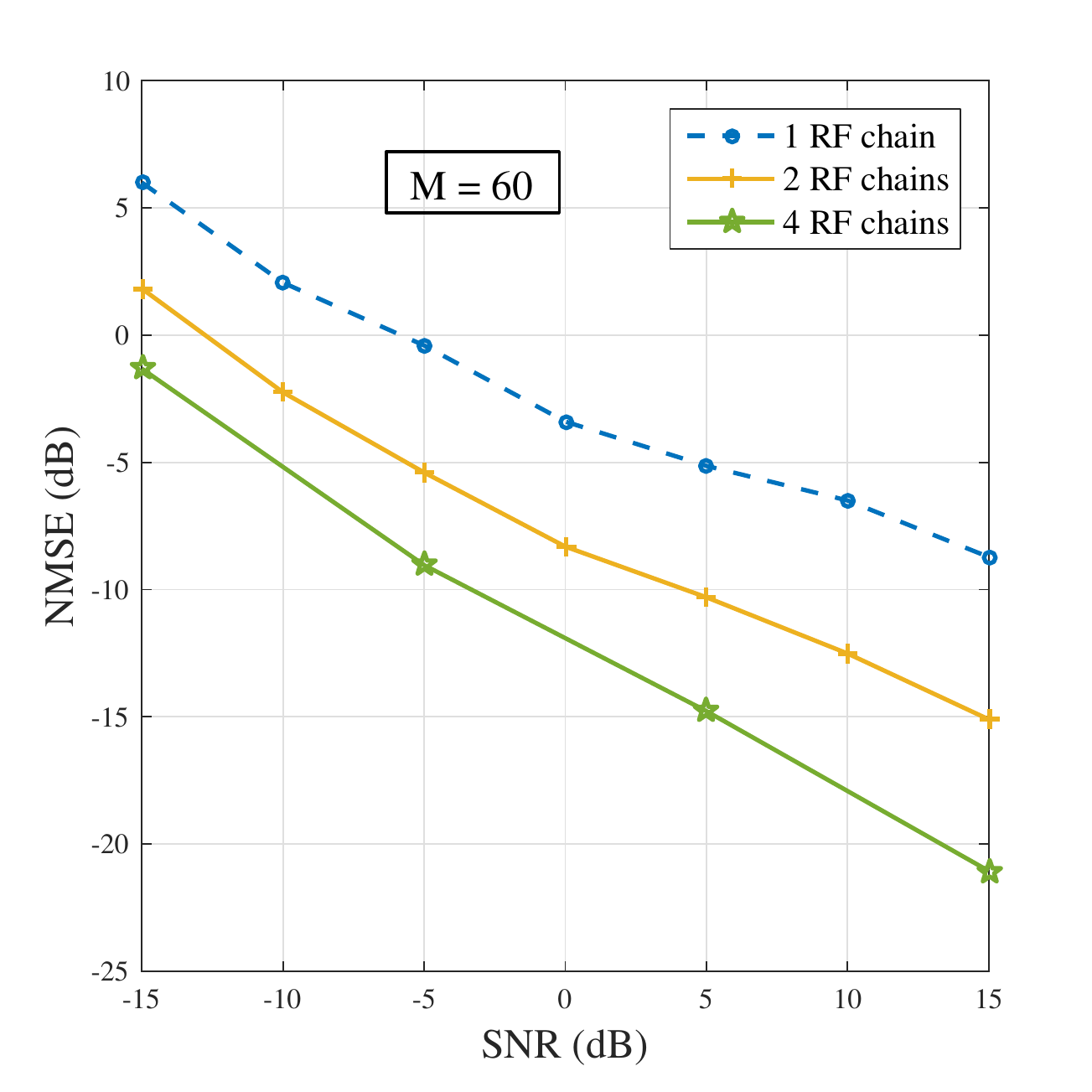} 
	\caption{Average $\ttNMSE$ for the proposed time-domain channel estimation approach as a function of SNR for different numbers of RF chains used at the transceivers. By employing multiple RF chains at the transceivers, the $\ttNMSE$ performance is improved.}     
	\label{fig:NMSEwithComb}        
\end{figure}
}{
\begin{figure}
	\centering
	\includegraphics[width=5in,height=4in]{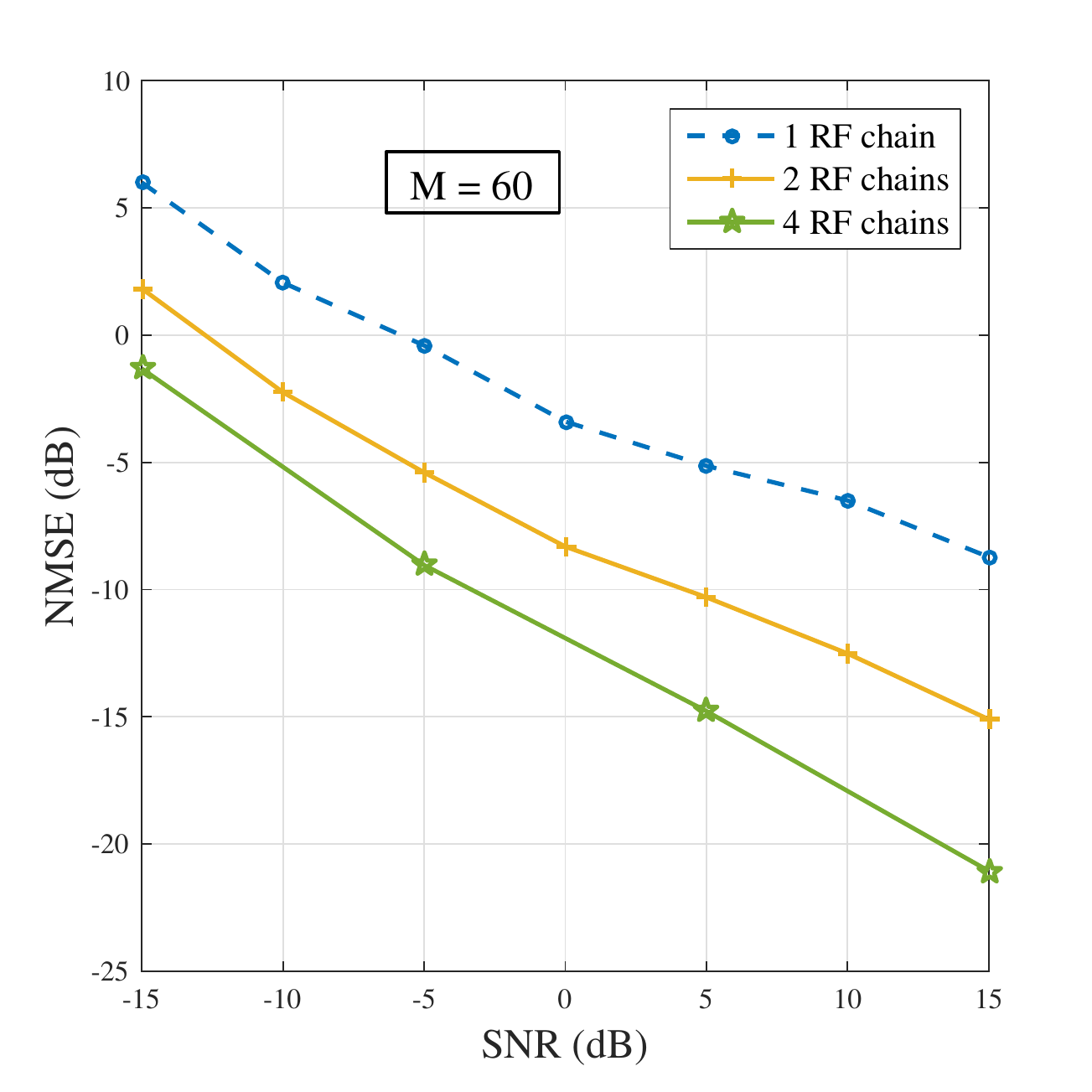} 
	\caption{Average $\ttNMSE$ for the proposed time-domain channel estimation approach as a function of SNR for different numbers of RF chains used at the transceivers. By employing multiple RF chains at the transceivers, the $\ttNMSE$ performance is improved.}     
	\label{fig:NMSEwithComb}        
\end{figure}}

Fig. \ref{fig:NMSEwithMeas} shows the $\ttNMSE$ as a function of the post combining received signal SNR using the proposed time-domain channel estimation approach.  Here we assume $\Nr = 16$, $\Nt = 32$, $\Nc = 4$, $N = 16$ and $\Np=2$. The time domain dictionary is constructed with the parameters $\Gr = 32$, $\Gt = 64$, and $\Gc = 8$. From Fig. \ref{fig:NMSEwithMeas}, it can be seen that with training length of even $80-100$ frames, sufficiently low channel estimation error can be ensured. For comparing the impact of angle quantization error, we show the $\ttNMSE$ for the case when the AoAs/AoDs of the mmWave channel are drawn from quantized grids with $\Gt = 64$ and $\Gr = 32$ that are used to construct the dictionaries, and also the case when the AoAs/AoDs are unrestricted. Choosing larger values for $\Gr~ (\Gt)$ in comparison with $\Nr~ (\Nt)$ can further narrow the error gap between the two cases, so does increasing $\Gc$ in comparison to $\Nc$ as the dictionary will become more and more \textit{robust}.

\iftoggle{2column}{
\begin{figure}
	\centering
	\includegraphics[width=3.25in,height=2.6in]{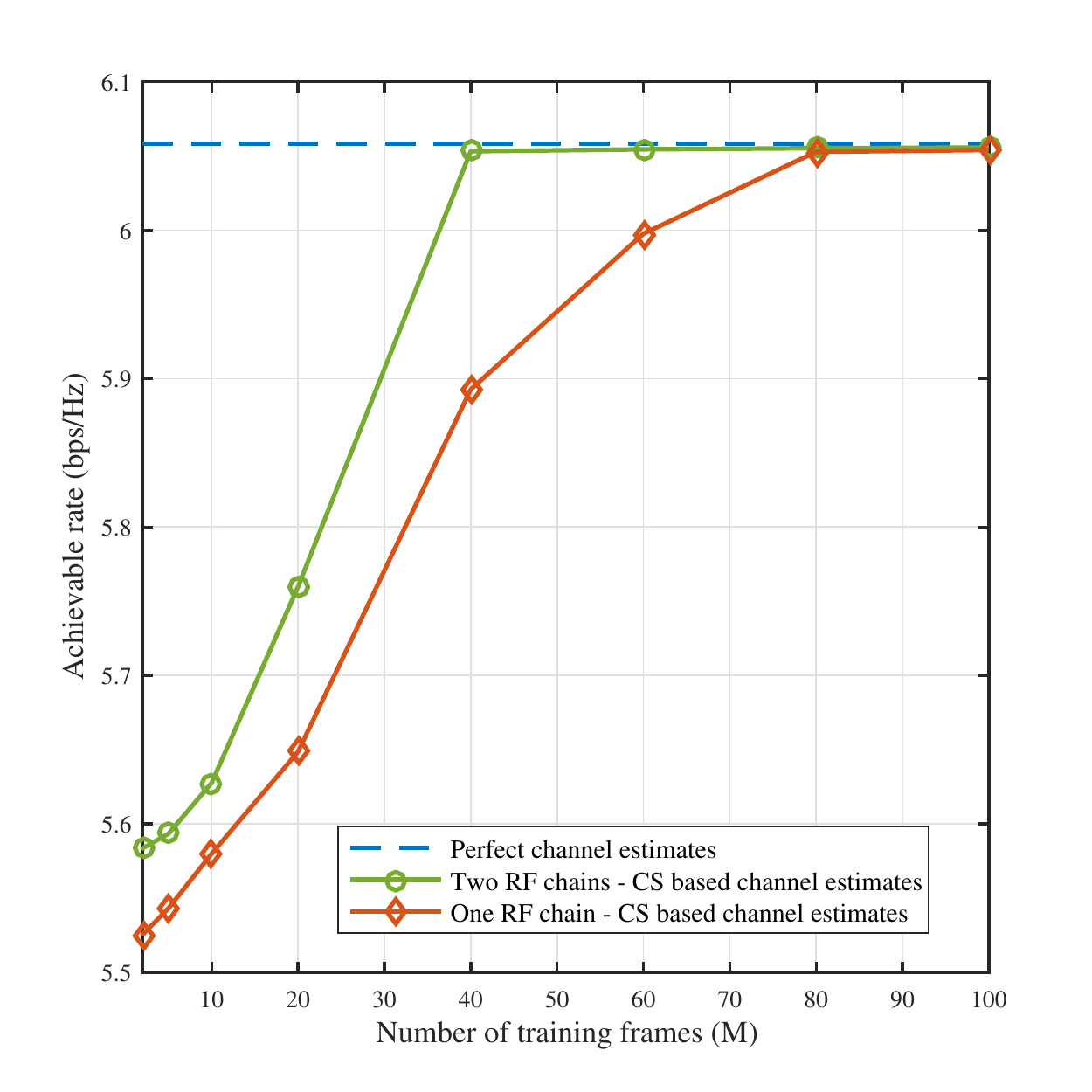} 
	\caption{Achievable spectral efficiency using the proposed time-domain channel estimation approach as a function of the number of training frames used $M$ for different numbers of RF combiners $\NRF$ used at the receiver. Employing multiple RF chains at the transceivers significantly reduces the number of training steps.}     
	\label{fig:Spec_effwithMNm}        
\end{figure}
}{
\begin{figure}
	\centering
	\includegraphics[width=5in,height=4in]{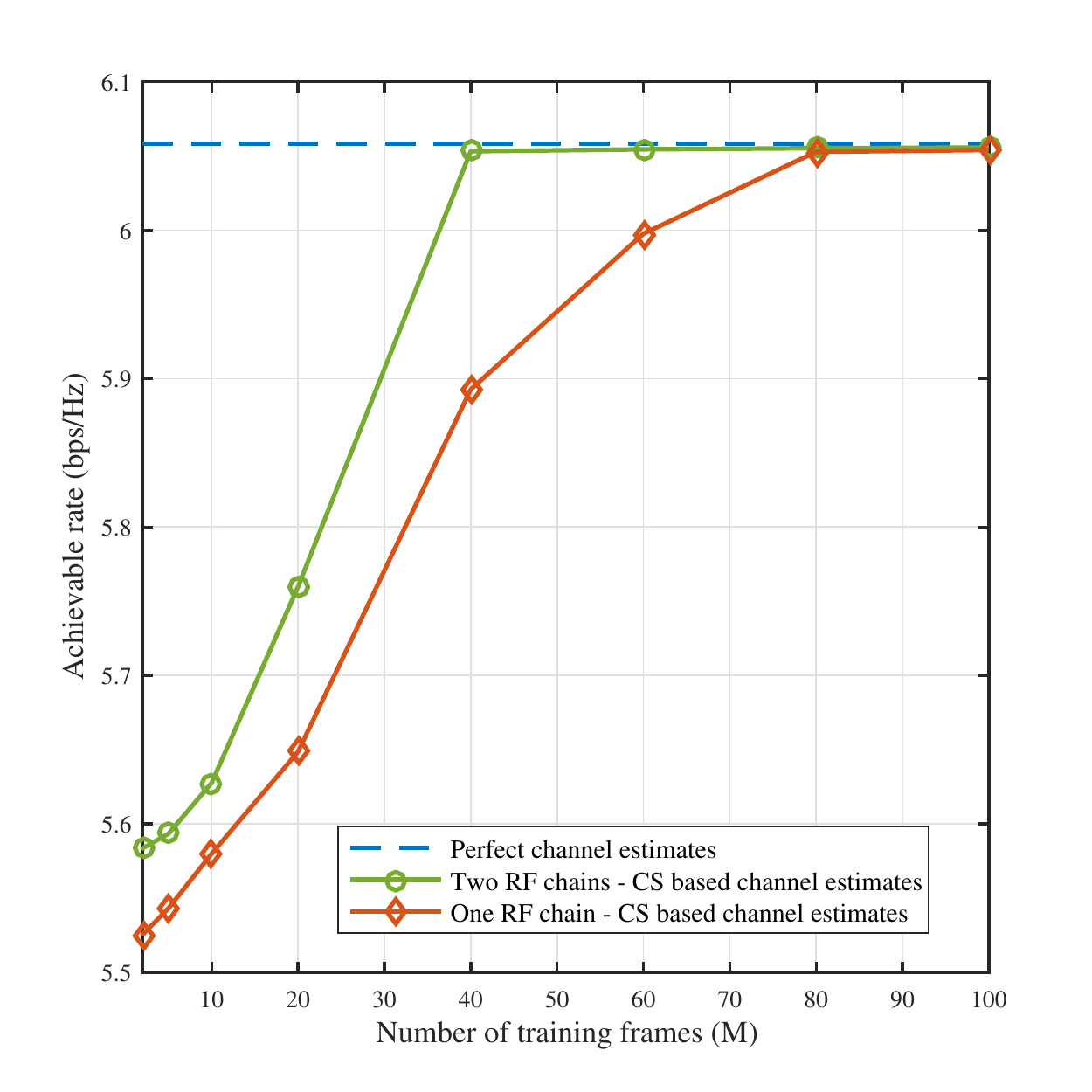} 
	\caption{Achievable spectral efficiency using the proposed time-domain channel estimation approach as a function of the number of training frames used $M$ for different numbers of RF combiners $\NRF$ used at the receiver. Employing multiple RF chains at the transceivers significantly reduces the number of training steps.}     
	\label{fig:Spec_effwithMNm}        
\end{figure}
}

\figref{fig:NMSEwithComb} shows how employing multiple RF chains at the transmitter and receiver can give good improvement in the estimation performance while requiring fewer number of training frame transmissions. We assume the same set of parameters as those used for generating \figref{fig:NMSEwithMeas}, and the proposed time-domain channel estimation approach, while altering the number of RF chains used at the transceivers. In \figref{fig:NMSEwithComb}, we assume $M=60$ frames are transmitted for training. The improvement in $\ttNMSE$ performance occurs thanks to a larger number of effective combining beam patterns that scale with the number of RF combiners $\NRF$ at the receiver. Similarly, employing multiple RF chains $\NRF$ at the transmitter contributes to a larger set of random precoders, resulting in smaller estimation error via compressed sensing. So, larger $\NRF$ is preferred to decrease the estimation error and to fully leverage the hybrid architecture in wideband mmWave systems.

In both  \figref{fig:NMSEwithMeas} and \figref{fig:NMSEwithComb}, we considered averaged $\ttNMSE$ to highlight the effectiveness of the proposed time-domain channel estimation algorithm and the performance gain when multiple RF chains are employed at the transmitter and receiver. In Fig. \ref{fig:Spec_effwithMNm}, the achievable spectral efficiency is plotted as a function of number of training steps $M$. We assume the same set of parameters as that in \figref{fig:NMSEwithMeas}. It is observed that having more RF combiners results in fewer number of training frame transmissions to achieve the same spectral efficiency. This is because, with multiple RF chains at the receiver, more effective measurements can be obtained per training frame that is transmitted.

\iftoggle{2column}{
\begin{figure}
	\centering
	\includegraphics[width=3.25in,height=2.6in]{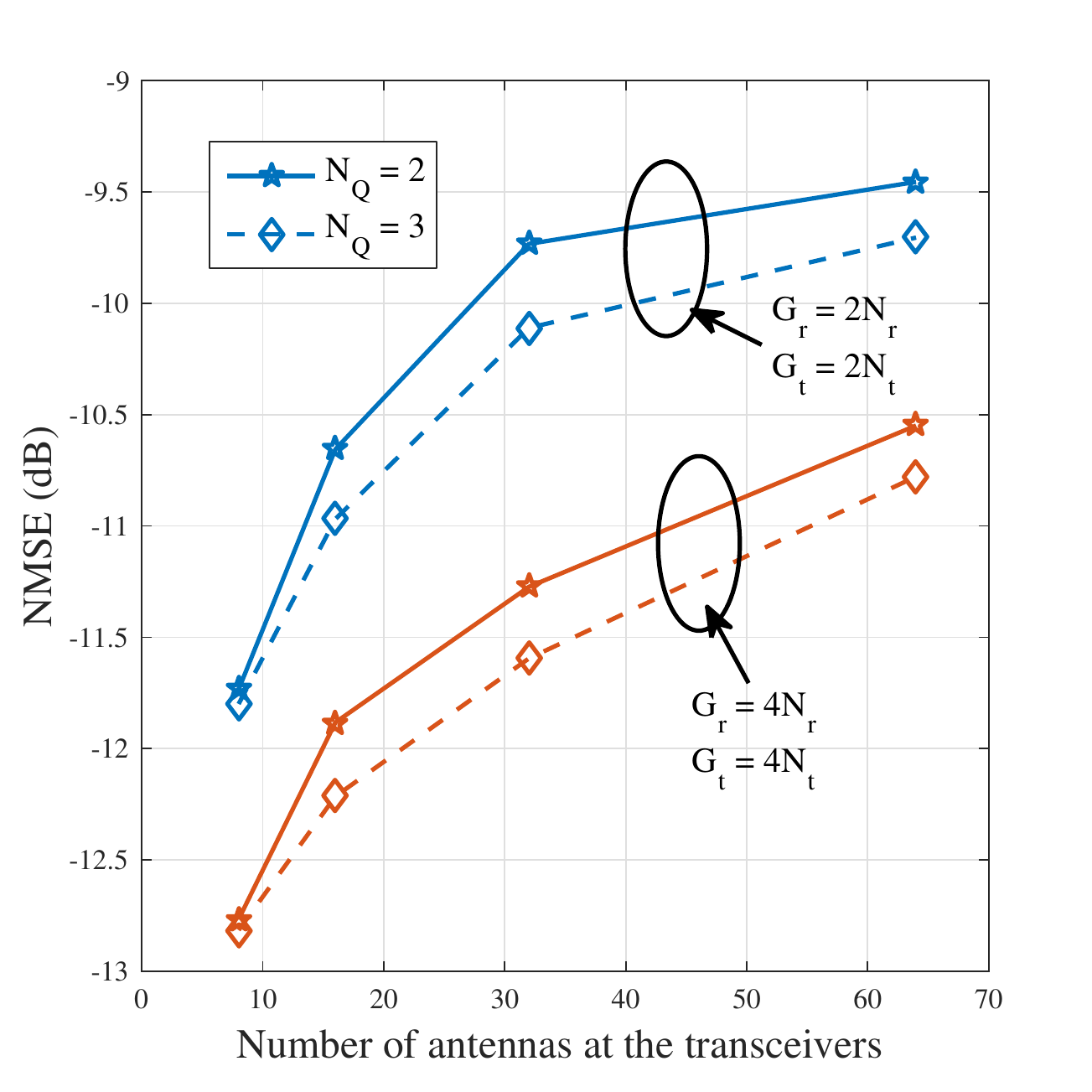} 
	\caption{Average $\ttNMSE$ versus the number of antenna elements (assuming $\Nr = \Nt$) using the proposed frequency-domain channel estimation approach. The number of angles in the quantized grid used for generating the dictionary is denoted as $\Gr$ (for AoA) and $\Gt$ (for AoD). The figure shows plots for different number of bits $\NQ$ used for angle quantization in the phase shifters during the training phase. }     
	\label{fig:NMSEwithAntenna_fd}        
\end{figure}
}{
\begin{figure}
	\centering
	\includegraphics[width=5in,height=4in]{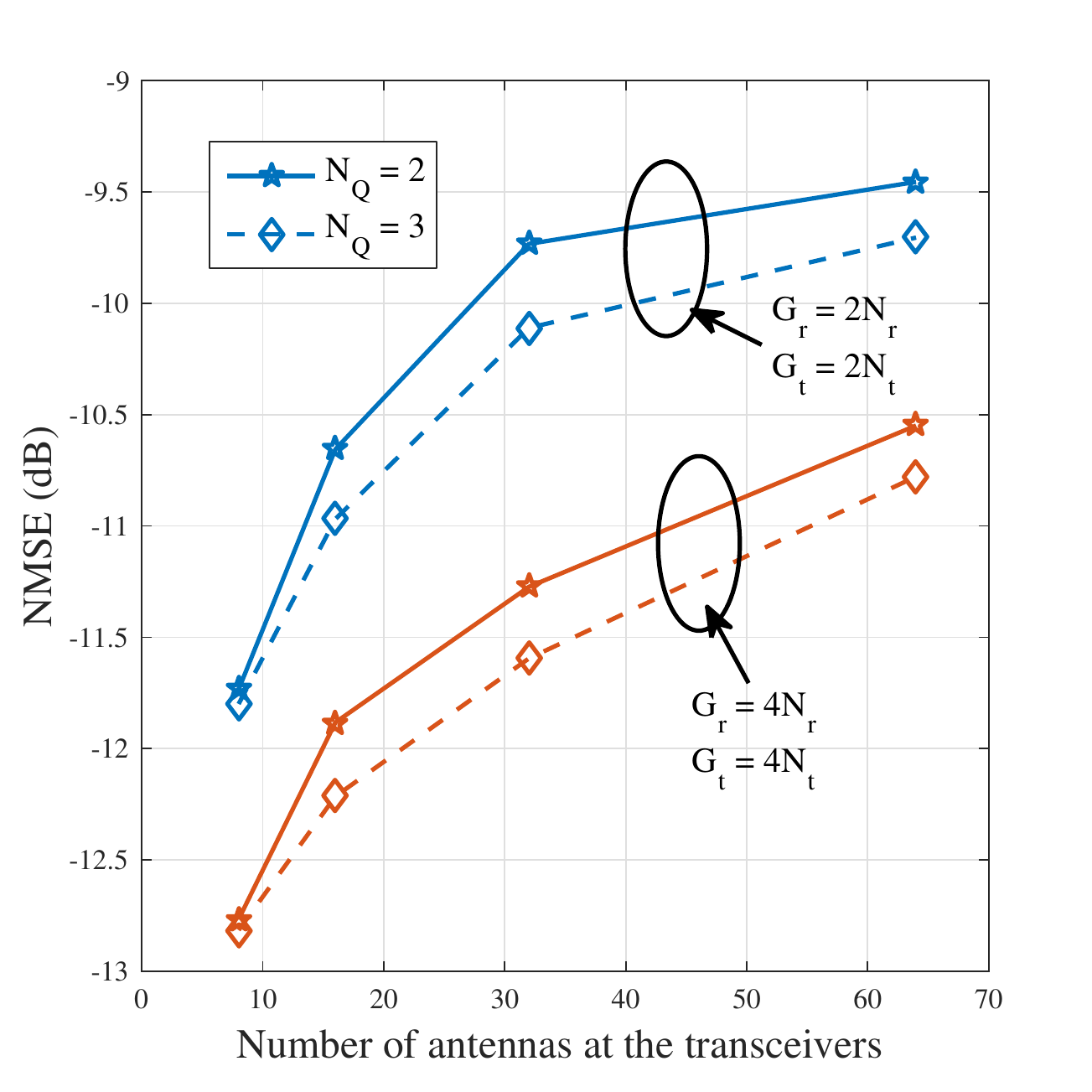} 
	\caption{Average $\ttNMSE$ versus the number of antenna elements (assuming $\Nr = \Nt$) using the proposed frequency-domain channel estimation approach. The number of angles in the quantized grid used for generating the dictionary is denoted as $\Gr$ (for AoA) and $\Gt$ (for AoD). The figure shows plots for different number of bits $\NQ$ used for angle quantization in the phase shifters during the training phase. }     
	\label{fig:NMSEwithAntenna_fd}        
\end{figure}
}

\iftoggle{2column}{
\begin{figure}
	\centering
	\includegraphics[width=3.25in,height=2.6in]{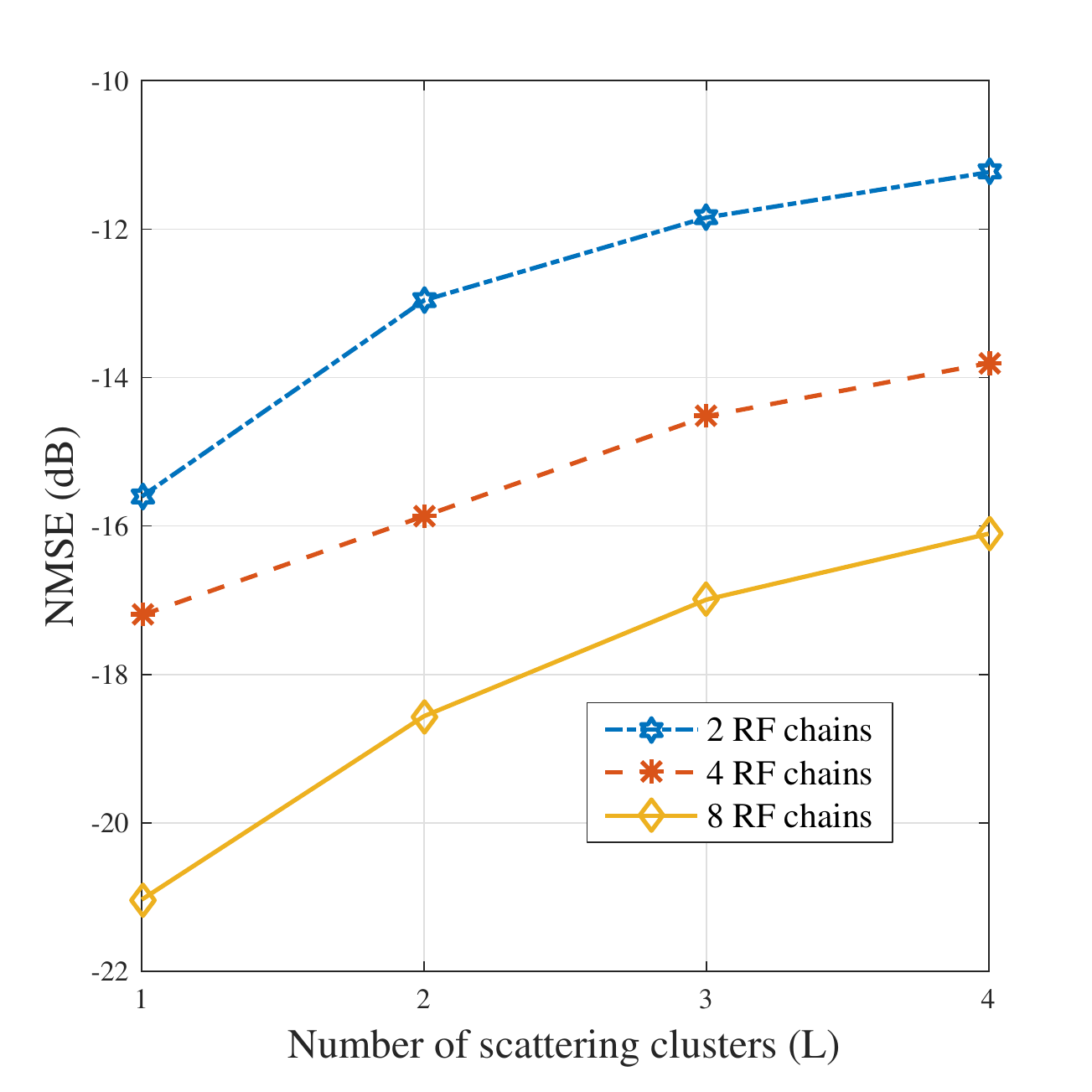} 
	\caption{Average $\ttNMSE$ versus the number of paths $\Np$, for different hybrid configurations at the transceivers using the proposed combined time-frequency domain channel estimation approach. Increasing $\Np$ increases the number of unknown parameters of the channel, and hence higher number of compressive measurements are required to get the required target estimation error performance.}     
	\label{fig:NMSEwithClusters_joint}        
\end{figure}
}{
\begin{figure}
	\centering
	\includegraphics[width=5in,height=4in]{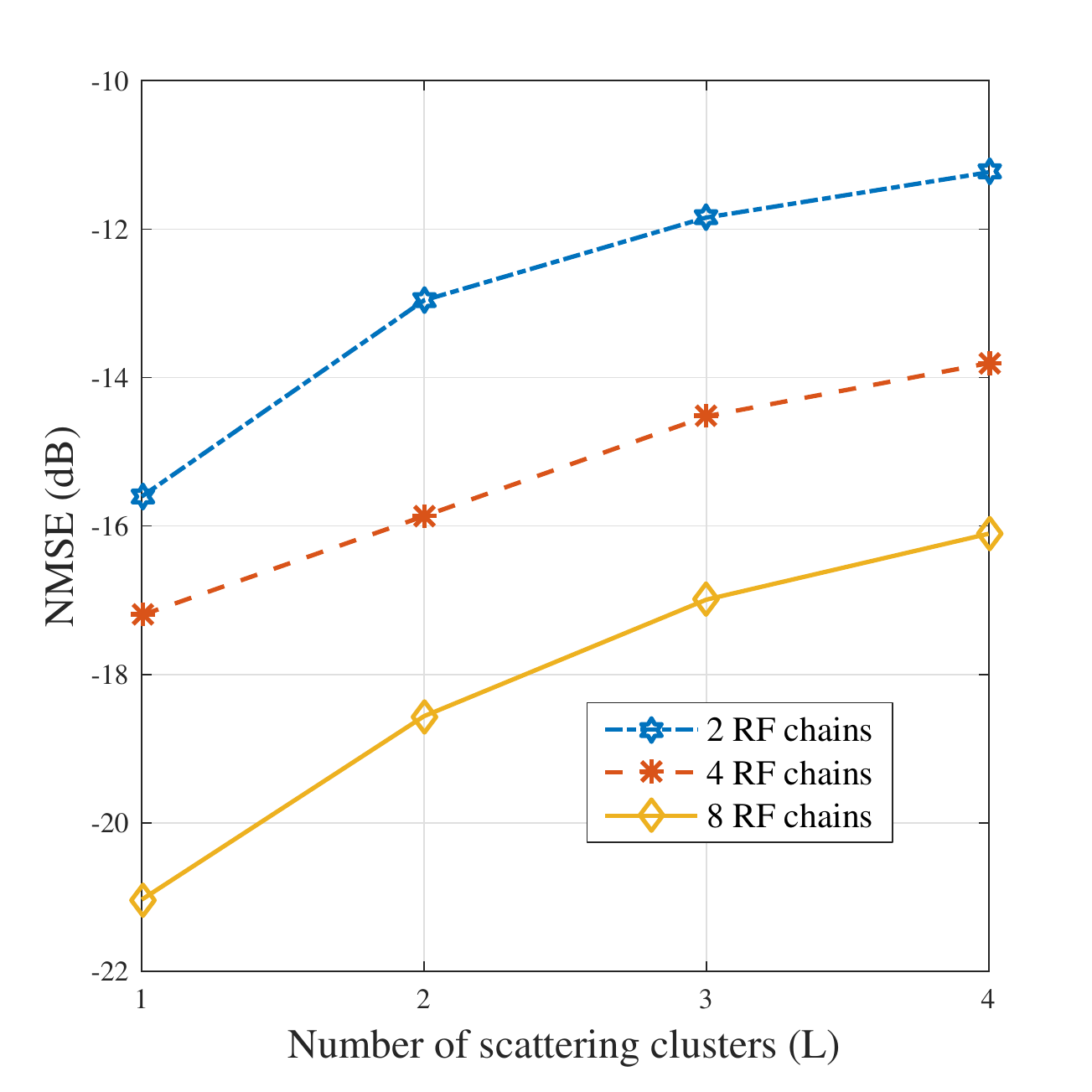} 
	\caption{Average $\ttNMSE$ versus the number of paths $\Np$, for different hybrid configurations at the transceivers using the proposed combined time-frequency domain channel estimation approach. Increasing $\Np$ increases the number of unknown parameters of the channel, and hence higher number of compressive measurements are required to get the required target estimation error performance.}     
	\label{fig:NMSEwithClusters_joint}        
\end{figure}
}

In \figref{fig:NMSEwithAntenna_fd}, we study the performance of the proposed frequency-domain channel estimation approach. We assume the number of compressive estimation training steps $M = 60$, the frame length $N = 16$, $\NRF = 2$ RF chains at the transceivers, and the number of delay taps $\Nc = 4$. The number of paths $\Np$ is assumed to be 2. The size of the FFT block used is $K = N = 16$. \figref{fig:NMSEwithAntenna_fd} shows the $\ttNMSE$ as a function of the number of antenna elements at the transceivers, with $\Nr = \Nt$ for different values of the dictionary parameters and number quantization bits used for generating the random phases in the precoders and combiners. While $\ttNMSE$ increases with increase in the size of the antenna array due to more ambiguity happening in the array response vector, increasing the size of the angle grid in comparison to the antenna size, improves the $\ttNMSE$. Increasing $\NQ$, the number of bits used in the measurement matrix, improves the efficiency of the compressive measurements by contributing more randomness. Thus a higher $\NQ$ is good from robust estimation point of view, though it results in more feedback overhead bits in the system.

In \figref{fig:NMSEwithClusters_joint}, we plot the $\ttNMSE$ as a function of the number of paths in the channel for various RF combiner setups at the receiver. The proposed combined time-frequency domain channel estimation approach is used with $\Nr = \Nt = 32$, $\Gr = \Gt = 64$, $N = 32$, $\Nc=8$ and $M = 60$ compressive training steps. In \figref{fig:NMSEwithClusters_joint}, we use sparse recovery in $P=1$ subcarrier in the frequency domain to estimate the AoAs and AoDs, before switching to the time domain to estimate all the gain coefficients of the channel paths. As $\Np$ is increased, the number of unknown parameters in the channel increases, thus increasing the estimation error for a given number of training steps and hardware configuration. Increasing the RF combiners, however, helps reduce the $\ttNMSE$ to meet a target estimation error performance. 

\iftoggle{2column}{
\begin{figure}
	\centering
	\includegraphics[width=3.25in,height=2.6in]{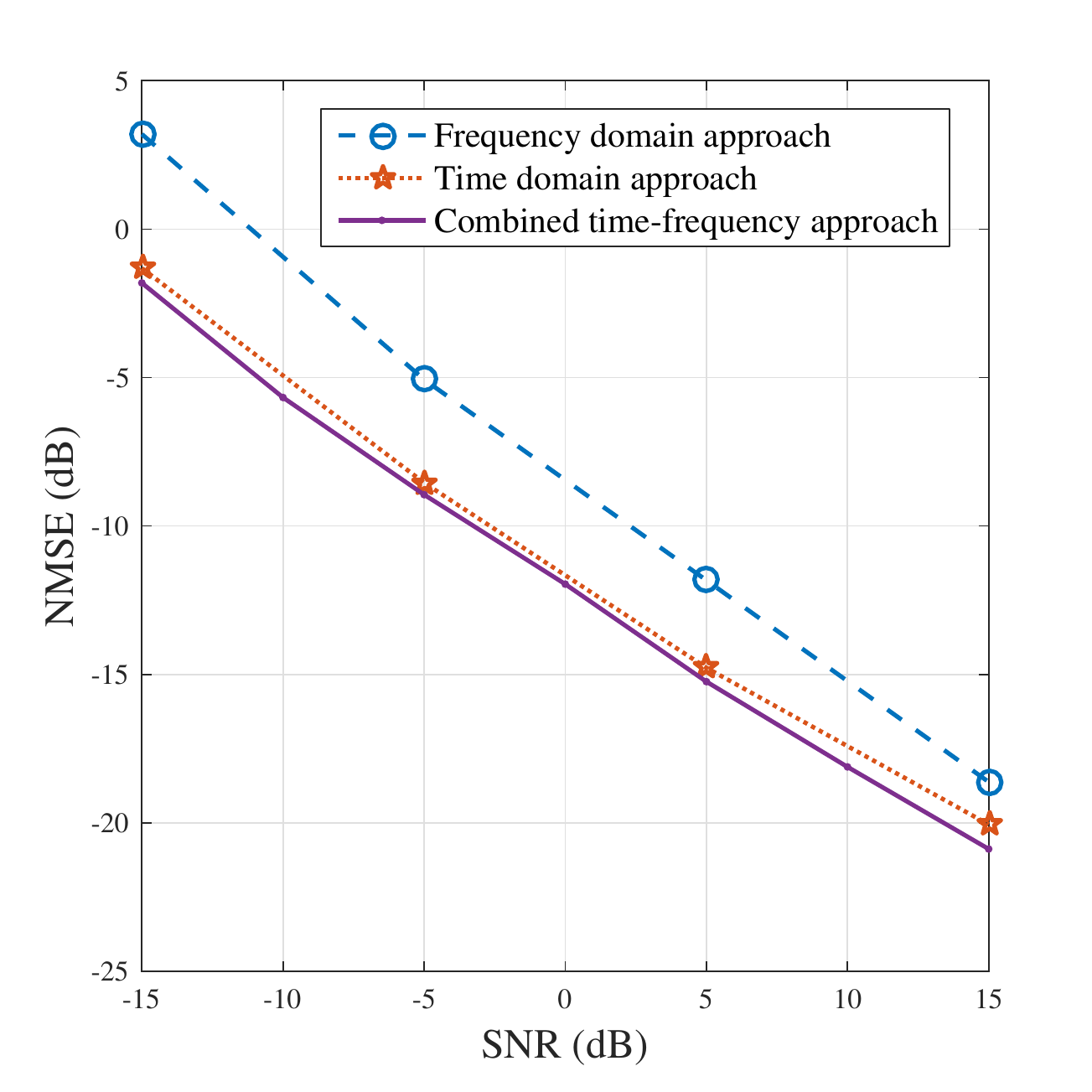} 
\caption{Plot showing the error performance of the three compressed sensing based channel estimation approaches proposed in the paper as a function of SNR. At low $\SNR$ the combine time-frequency approach has the least average $\ttNMSE$, while at higher $\SNR$s, all the three proposed approaches give similar performance. }
\label{fig:NMSEdiffApproaches}        
\end{figure}
}{
\begin{figure}
	\centering
\includegraphics[width=5in,height=4in]{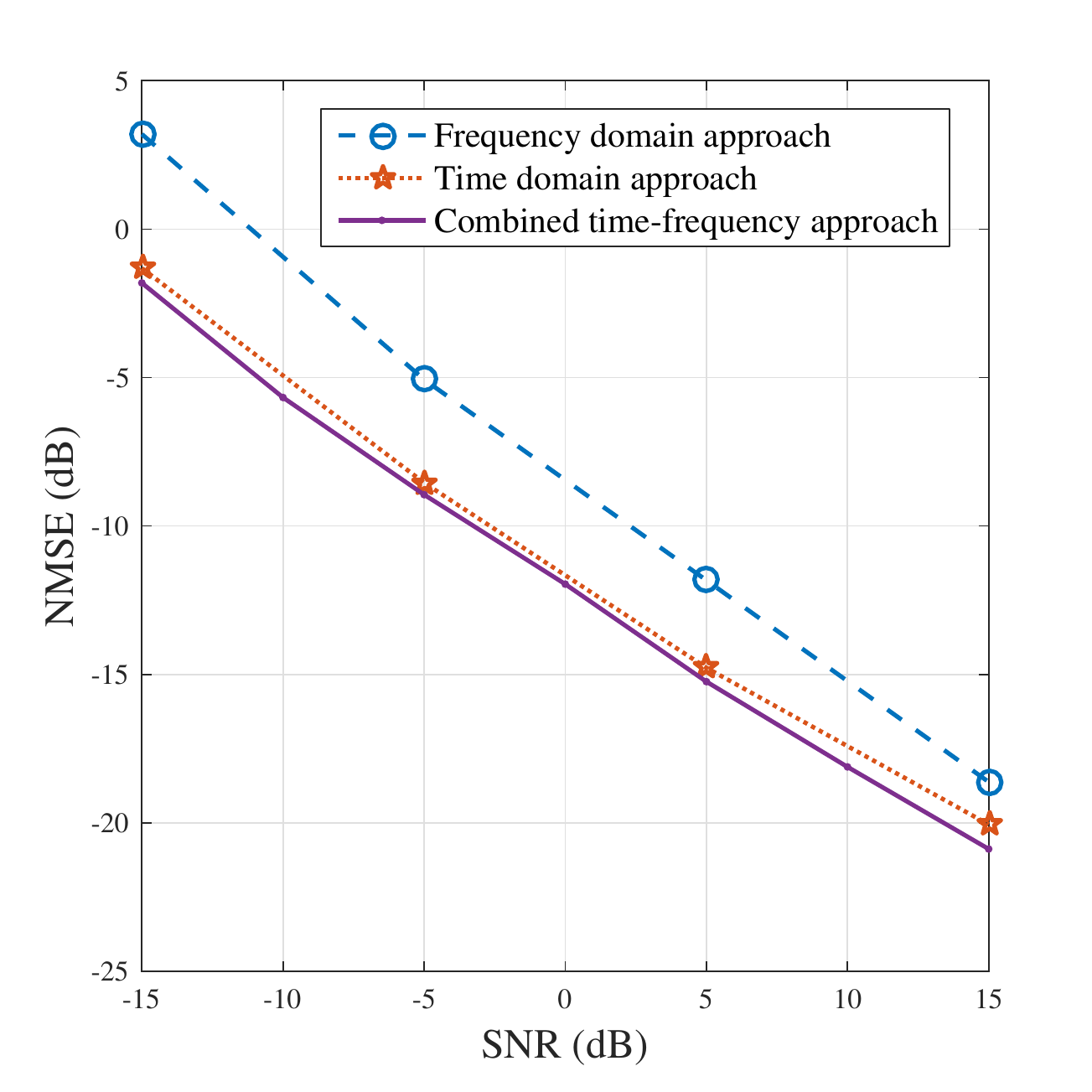} 
\caption{Plot showing the error performance of the three compressed sensing based channel estimation approaches proposed in the paper as a function of SNR. At low $\SNR$ the combine time-frequency approach has the least average $\ttNMSE$, while at higher $\SNR$s, all the three proposed approaches give similar performance.}
\label{fig:NMSEdiffApproaches}        
\end{figure}
}
In \figref{fig:NMSEdiffApproaches}, we look the error performance of the three proposed channel estimation approaches by plotting the $\ttNMSE$ as a function of SNR. We assume $\Nr = \Nt = 32$, $\Gr = \Gt = 64$, $M = 60$, $\Nc = 4$ and $\Np = 2$. The number of RF chains at the transceivers $\NRF$ is assumed to be 4. The combined time-frequency approach is assumed to use OMP on $P=1$ subcarrier in the frequency domain to recover the angles of arrival and departure. For constructing the time domain dictionary, we assume $\Gc = 2\Nc$ delay quantization parameter. It can be seen that the combined time-frequency gives the best error performance whereas the proposed frequency domain approach results is large estimation error, especially at lower SNRs. This is mainly due to the accumulation of error incurred due to $K$ parallel OMPs in the frequency, which is avoided in the combined time-frequency approach and the proposed time domain approach, which invoke the sparse recovery algorithm only once (when $P=1$). At higher SNRs, however, the three proposed approaches give similar estimation error performance. 

\iftoggle{2column}{
\begin{figure}
	\centering
	\includegraphics[width=3.25in,height=2.6in]{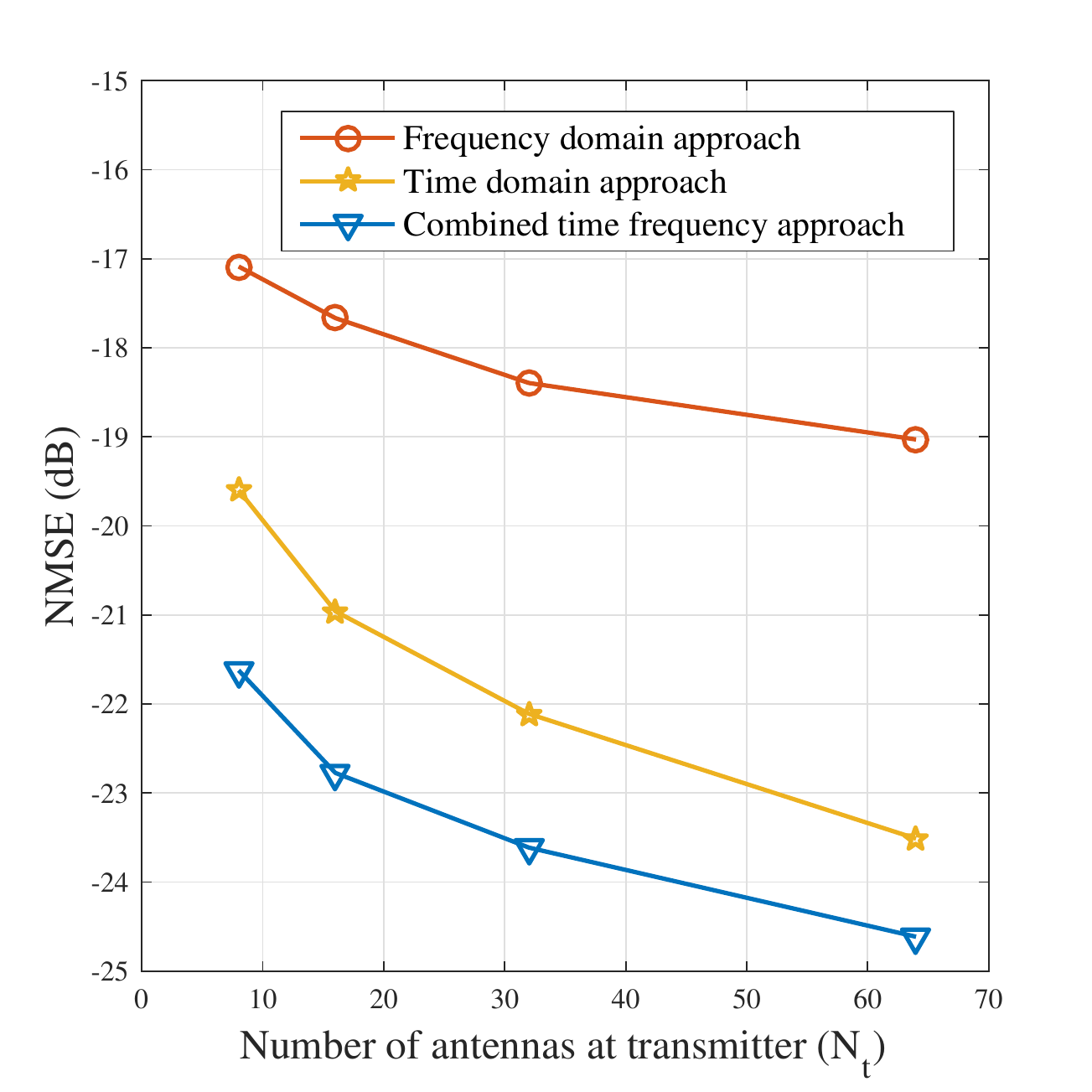} 
	\caption{Average $\ttNMSE$ performance with different number of antenna elements at the transmitter for the three proposed channel estimation approaches. For a given number of training length and angle quantization grid size large enough, larger number of antennas leads to better error performance in all the three proposed approaches.}
	\label{fig:NMSEwithNrNt_all}        
\end{figure}
}{
\begin{figure}
	\centering
	\includegraphics[width=5in,height=4in]{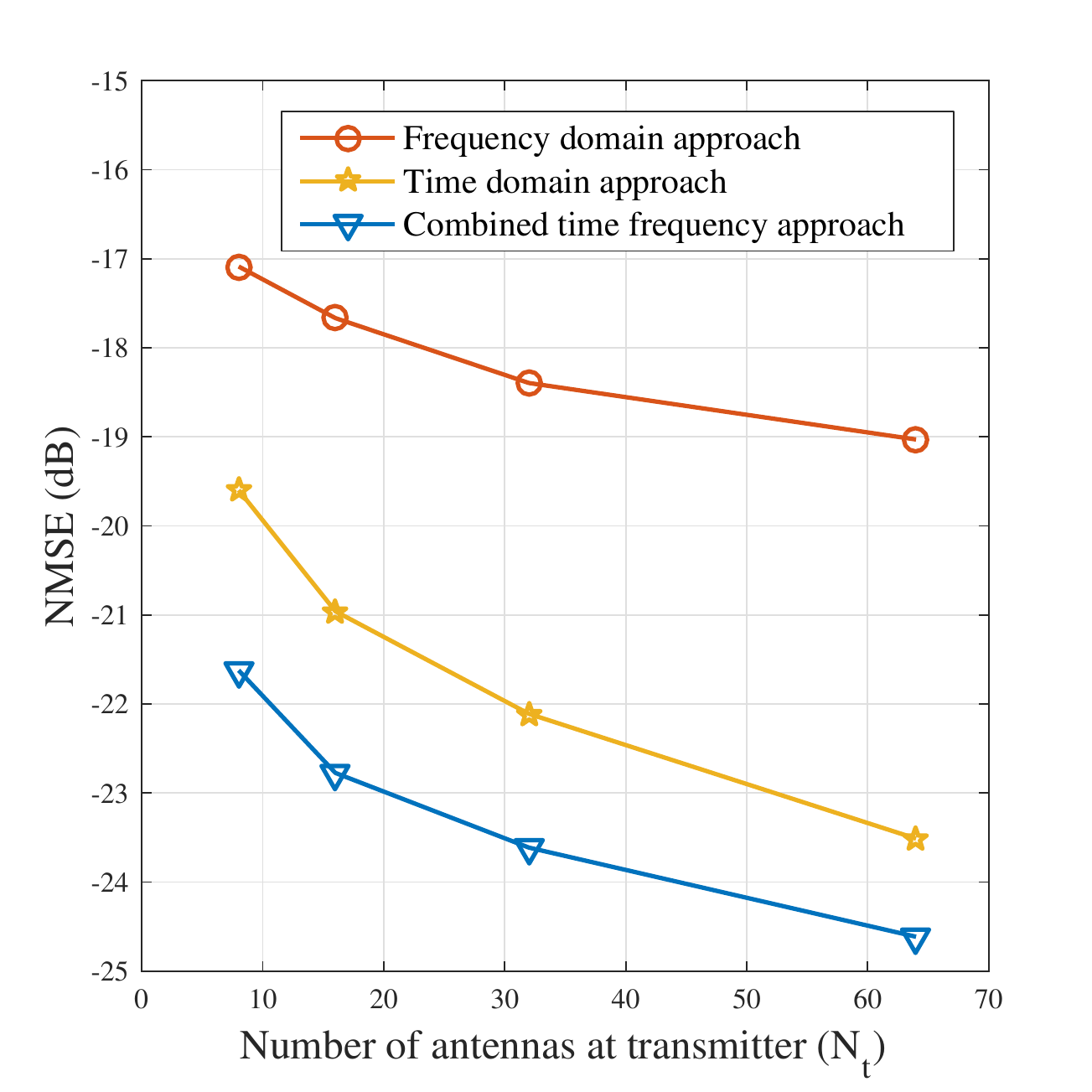} 
	\caption{Average $\ttNMSE$ performance with different number of antenna elements at the transmitter for the three proposed channel estimation approaches. For a given number of training length and angle quantization grid size large enough, larger number of antennas leads to better error performance in all the three proposed approaches.}
	\label{fig:NMSEwithNrNt_all}        
\end{figure}
}

In \figref{fig:NMSEwithNrNt_all}, we compare the error performance of the three proposed channel estimation approaches as a function of the number of antennas used at the transmitter. We let $\Gr = 2\Nr = 16$ and $\Gt = 100$ here and assume the AoA and AoD to fall on these grid values, for constructing the dictionaries. The length of training frames is fixed to $N=16$ for a total of $M = 60$ training steps, and 4 RF chains are assumed at the transceivers. In a frequency selective mmWave channel with $\Nc = 4$ delay taps and $\Np=2$ paths, \figref{fig:NMSEwithNrNt_all} shows that increasing the antenna array results in decreased estimation error when the quantization grid size is large enough that the actual angles of arrival and departure fall in the grid. We assumed the number of quantized angles used for the phase shifters at the transceivers does not scale with the number of antenna elements. The quantized angles were drawn uniformly at random from the set $\{1,~ \jj,~ -1,~ -\jj \}$.

\iftoggle{2column}{
\begin{figure}
	\centering
	\includegraphics[width=3.25in,height=2.6in]{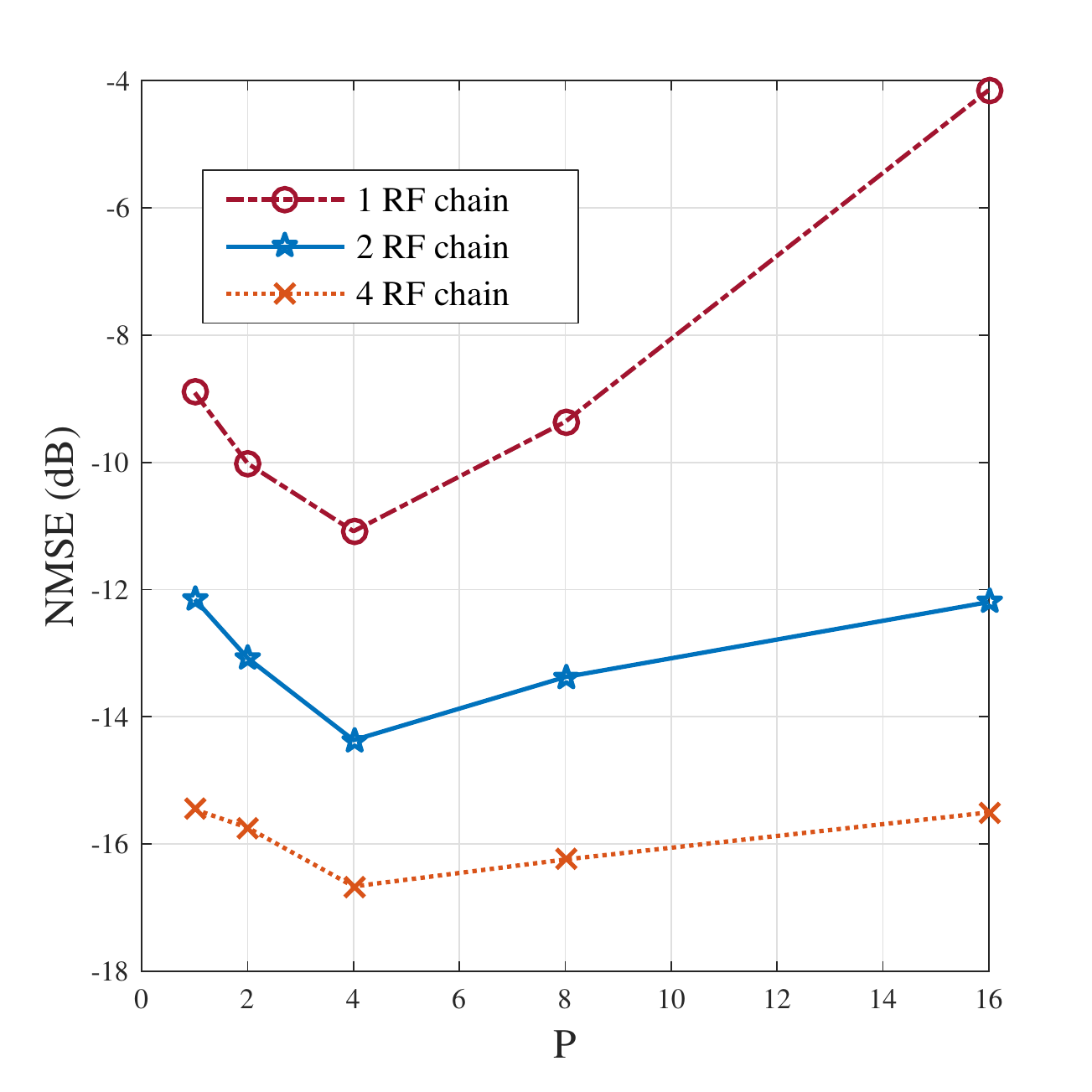} 
\caption{Average $\ttNMSE$ as a function of $P$, the number of subcarriers in the frequency domain used for independent OMP based AoA/AoD support recovery, for different hardware configurations using the combined time-frequency compressive channel estimation.}
\label{fig:NMSEwithP_joint}        
\end{figure}
}{
\begin{figure}
	\centering
\includegraphics[width=5in,height=4in]{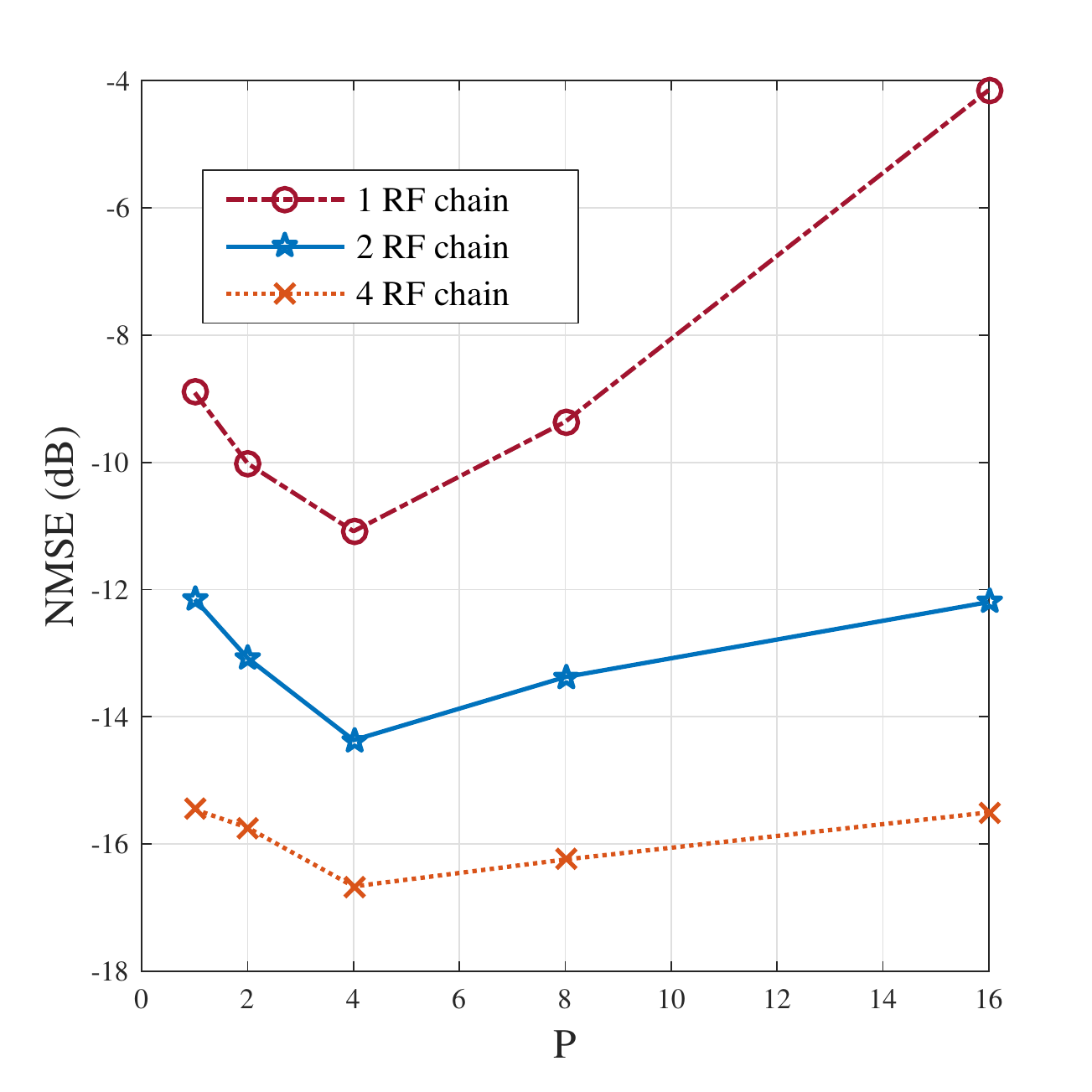} 
\caption{Average $\ttNMSE$ as a function of $P$, the number of subcarriers in the frequency domain used for independent OMP based AoA/AoD support recovery, for different hardware configurations using the combined time-frequency compressive channel estimation.}
\label{fig:NMSEwithP_joint}        
\end{figure}
}
The choice of $P$, the number of subcarriers used in the frequency domain to perform parallel and independent OMP based AoA/AoD estimation is important. In \figref{fig:NMSEwithP_joint}, we plot $\ttNMSE$ as a function of $P$ for the proposed combined time-frequency domain channel estimation approach. Though we vary the hardware configuration at the receiver, it can be seen in \figref{fig:NMSEwithP_joint} that $P=4$ results in the best estimation error performance.

In \figref{fig:NMSEwithL_all}, we compare the three proposed approaches' error performance as the number of channel paths is increased. We set $\Gr = \Gt = 2\Nr = 2\Nt = 64$, and assume 4 RF chains at the transceivers. The training frame length of $N=16$ is assumed for the wideband channel of $\Nc=4$ delay taps. For each case, $M=60$ training steps are assumed. As the number of channel paths is increased, all the approaches perform worse. In particular, for a given delay quantization parameter $\Gc = 2\Nc$, assumed for the time domain approach's plot in \figref{fig:NMSEwithL_all}, the degradation in $\ttNMSE$ is significant as the delay estimation results in more error. For smaller number of channel paths, however, the time domain approach gives lower channel estimation error.
\iftoggle{2column}{
\begin{figure}
	\centering
	\includegraphics[width=3.25in,height=2.6in]{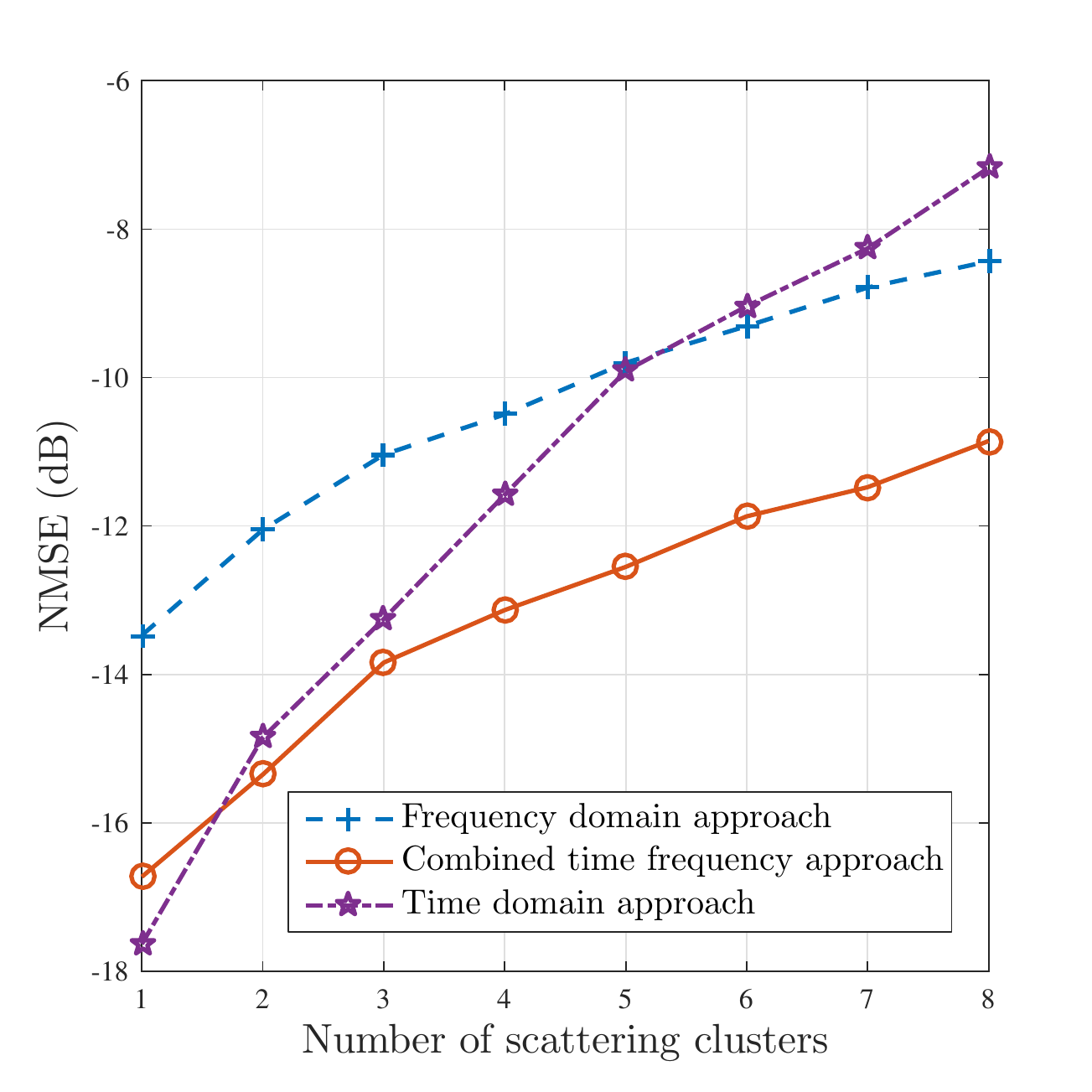} 
\caption{Plot showing the error performance of the three proposed approaches, as a function of the number of paths $\Np$ in the channel. Increasing $\Np$ degrades the average $\ttNMSE$ performance. While the proposed time domain approach gives the minimum average $\ttNMSE$ when the number of paths is small, the combine time-frequency approach gives the best error performance for larger $\Np$.}
\label{fig:NMSEwithL_all}        
\end{figure}
}{
\begin{figure}
	\centering
\includegraphics[width=5in,height=4in]{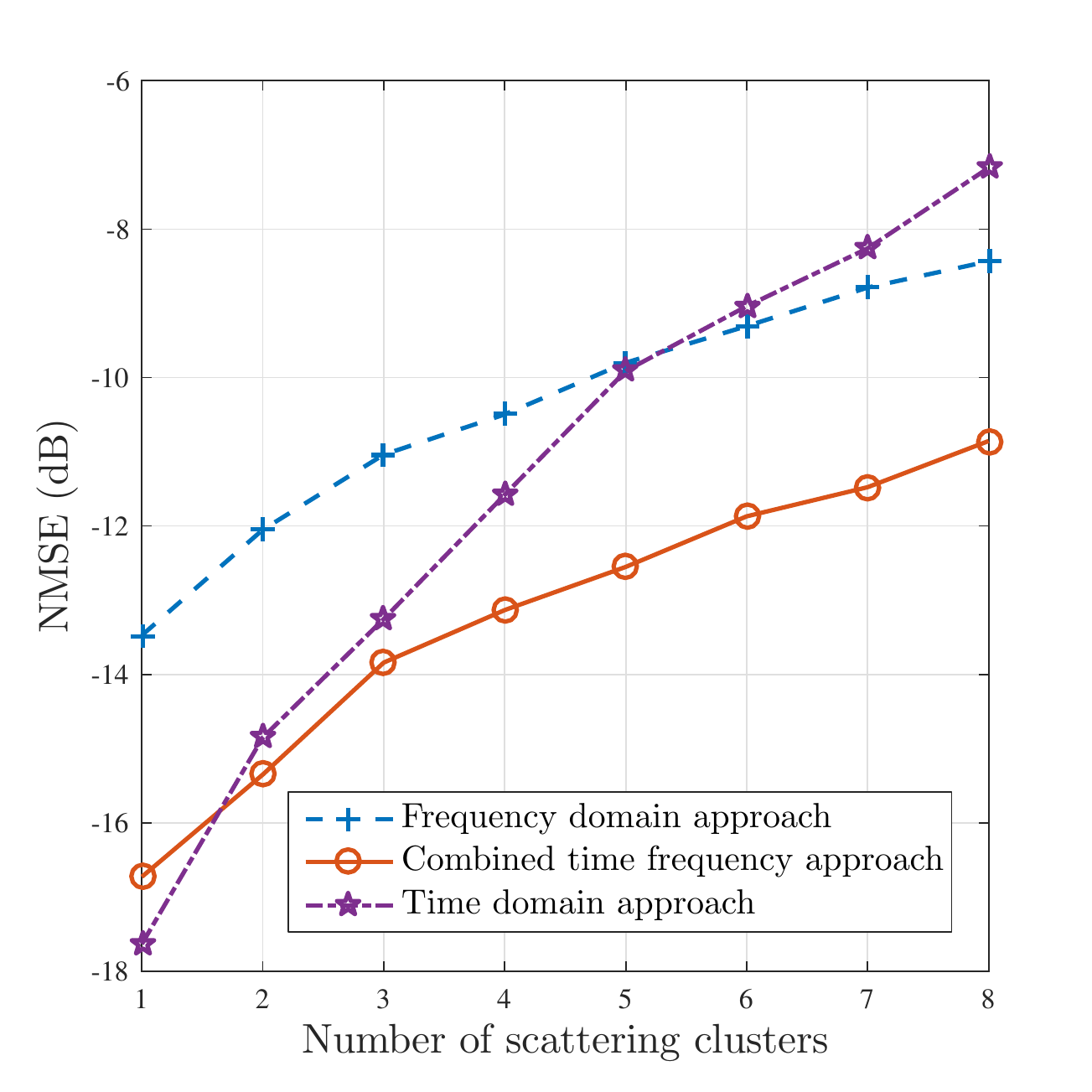} 
\caption{Plot showing the error performance of the three proposed approaches, as a function of the number of paths $\Np$ in the channel. Increasing $\Np$ degrades the average $\ttNMSE$ performance. While the proposed time domain approach gives the minimum average $\ttNMSE$ when the number of paths is small, the combine time-frequency approach gives the best error performance for larger $\Np$.}
\label{fig:NMSEwithL_all}        
\end{figure}
}

A comparison of the performance of the three proposed approaches as a function of the number of training steps is shown in \figref{fig:NMSEwithM_all}. We set the SNR to 5 dB here and assume $\Gr = \Gt = 2\Nr = 2\Nt = 64$. Each training frame is assumed to be of length 16 symbols, for a frequency selective mmWave channel of tap length 4, and channel paths 2. The $\ttNMSE$ plots in \figref{fig:NMSEwithM_all} is assumed 4 RF chains at the transceivers with 2 bit quantization at the phase shifters during channel estimation. It can be seen that while, with low training number of training steps the combined time frequency approach and the proposed frequency domain approach outperform the time domain approach, with larger number of training steps, the time domain approach gives the least $\ttNMSE$. 

To compare the overhead in channel training in the proposed compressive sensing based approaches, consider the short preamble structure used in IEEE 802.11ad \cite{IEEE:11ad}, which is of duration 1.891$\mu$s. At a chip rate of 1760 MHz, this short preamble consisting of the short training frame (STF) and the channel estimation frame (CEF) amounts to more than 3200 symbols. After the end of this short preamble transmission, IEEE 802.11ad beamforming protocol then switches to a different beam pair combination, and the process is repeated recursively to estimate the best set of beamforming directions. For the setting in \figref{fig:NMSEwithM_all}, however, $MN = 1600$ symbols are only required for the proposed approaches to achieve low average $\ttNMSE$ and explicit estimation of the frequency selective MIMO channel.
\iftoggle{2column}{
\begin{figure}
	\centering
	\includegraphics[width=3.25in,height=2.6in]{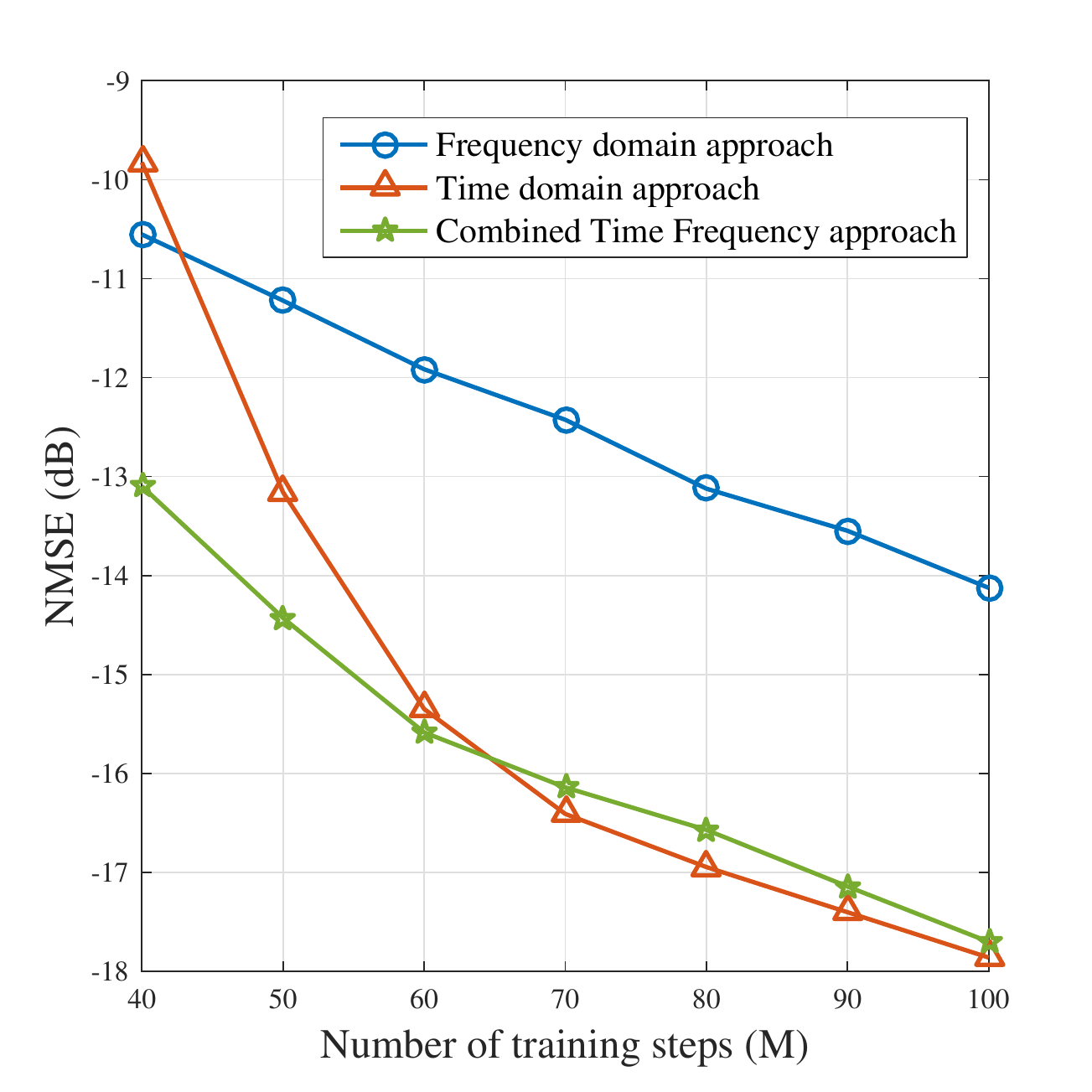} 
\caption{Plot showing the error performance of the three proposed approaches, as a function of the number of training steps $M$. More number of compressive measurements lead to better estimation error performance at the expense of higher signaling overhead. The combined time-frequency approach gives the best trade-off between low training overhead and minimum average $\ttNMSE$ performance.}
\label{fig:NMSEwithM_all}        
\end{figure}
}{
\begin{figure}
	\centering
\includegraphics[width=5in,height=4in]{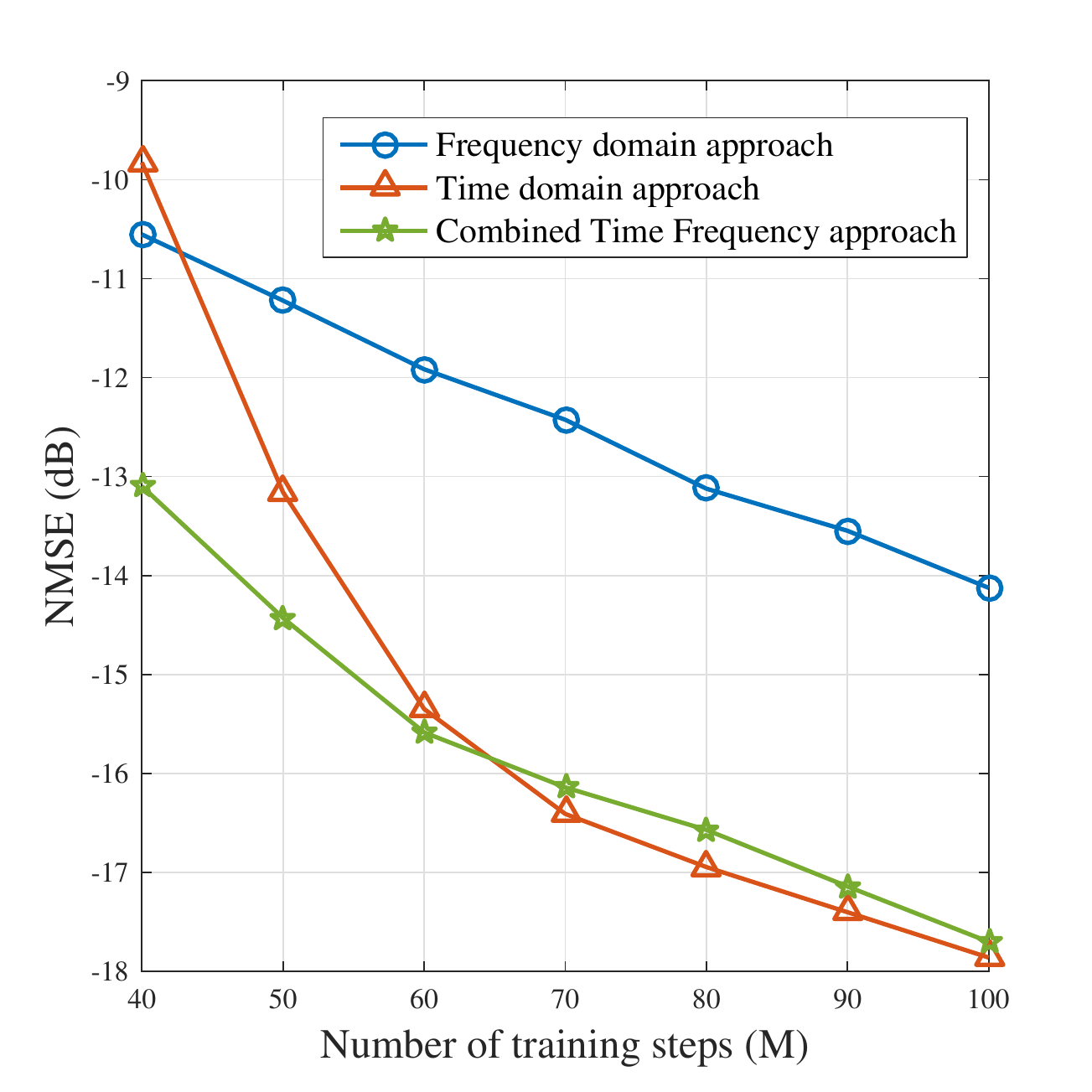} 
\caption{Plot showing the error performance of the three proposed approaches, as a function of the number of training steps $M$. More number of compressive measurements lead to better estimation error performance at the expense of higher signaling overhead. The combined time-frequency approach gives the best trade-off between low training overhead and minimum average $\ttNMSE$ performance.}
\label{fig:NMSEwithM_all}        
\end{figure}
}

\section{Conclusion}
In this paper, we proposed wideband channel estimation algorithms for frequency selective mmWave systems using a hybrid architecture at the transmitter and receiver. The system model adopts zero padding that allows enough time for switching the analog beams and, hence, well matches the hybrid architectures. The proposed channel estimation algorithms are based on sparse recovery and can support MIMO operation in mmWave systems since the entire channel is estimated after the beam training phase. Three different approaches - in purely time, in purely frequency and a combined time frequency approach were proposed, that can be used in both SC-FDE and OFDM based wideband mmWave systems.  Leveraging the frame structure and the hybrid architecture at the transceivers, it was shown that compressed sensing tools can be used for mmWave channel estimation. Simulation results showed that the proposed algorithms required very few training frames to ensure low estimation error. It was shown that further reduction in the training overhead and estimation error can be obtained by employing multiple RF chains at the transceivers.

\bibliographystyle{ieeetr}

\end{document}